\documentclass[twoside]{article}
\usepackage[norelsize]{algorithm2e}
\usepackage{url, fullpage}
\usepackage{rawfonts}
\usepackage{pgfplots}

\usepackage{color}

\usepackage{amsmath}
\usepackage{amsfonts}
\usepackage{amssymb}
\usepackage{comment}

\newcommand{\RR}{\mathbb{R}}


\begin{document}
\author{Jan Rupnik \and Andrej Muhi\v{c} \and Gregor Leban \and Primo\v{z} \v{S}kraba \and Bla\v{z} Fortuna \and Marko Grobelnik }
\title{News Across Languages - Cross-Lingual Document Similarity and Event Tracking}
%[Cross-Lingual Document Similarity and Event Tracking]

%
%\author{ Jan Rupnik } \email{jan.rupnik@ijs.si}
%\author{ Andrej Muhi\v{c} } \email{andrej.muhic@ijs.si}
%\author{ Gregor Leban } \email{gregor.leban@ijs.si}
%\author{ Primo\v{z} \v{S}kraba } \email{primoz.skraba@ijs.si}
%\author{ Bla\v{z} Fortuna } \email{blaz.fortuna@ijs.si}
%\author{ Marko Grobelnik } \email{marko.grobelnik@ijs.si}
%\address {Artificial Intelligence Laboratory, Jo\v{z}ef Stefan Institute,\\
%       Jamova cesta 39, 1000 Ljubljana, Slovenia}

\maketitle

\begin{abstract}

In today's world, we follow news which is distributed globally. Significant events are reported by different sources and in different languages. In this work, we address the problem of tracking of events in a large multilingual stream. Within a recently developed system Event Registry~\cite{Leban2014W,Leban2014I} we examine two aspects of this problem: how to compare articles in different languages and how to link collections of articles in different languages which refer to the same event.  Taking a multilingual stream and clusters of articles from each language, we compare different cross-lingual document similarity measures based on Wikipedia. This allows us to compute the similarity of any two articles regardless of language. Building on previous work, we show there are methods which scale well and can compute a meaningful similarity between articles from languages with little or no direct overlap in the training data.
Using this capability, we then propose an approach to link clusters of articles across languages which represent the same event. We provide an extensive evaluation of the system as a whole, as well as an evaluation of the quality and robustness of the similarity measure and the linking algorithm.
\end{abstract}

\section{Introduction}

Content on the Internet is becoming increasingly multilingual. A prime example is Wikipedia. In 2001, the majority of pages were written in English, while in 2015, the percentage of English articles has dropped to 14\%. At the same time, online news has begun to dominate reporting of current events. However, machine translation remains relatively rudimentary. It allows people to understand simple phrases on web pages, but remains inadequate for more advanced understanding of text. In this paper we consider the intersection of these developments: how to track events which are reported about in multiple languages.

\begin{figure}
\centering
\includegraphics[width=1\textwidth]{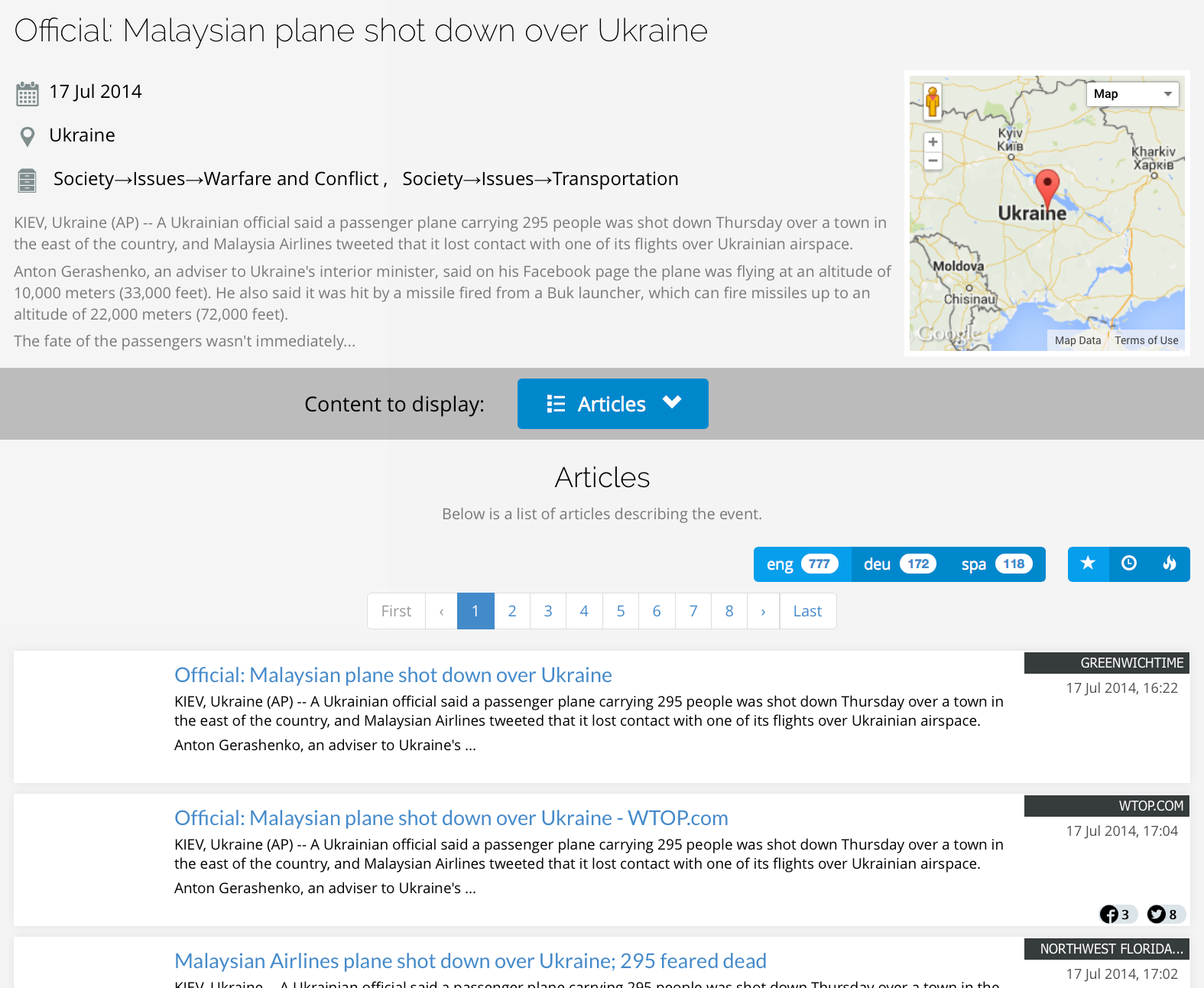}
\caption{\label{fig:event2} Events are represented by collections of articles about an event, in this case the Malaysian airliner which was shot down over Ukraine. The results shown in the figure can be obtained using the query \protect\url{http://eventregistry.org/event/997350\#?lang=eng\&tab=articles}. The content presented is part of the Event Registry system, developed by the authors.}
\end{figure}

The term event is vague and ambiguous, but for the practical purposes, we define it as ``any significant happening that is being reported about in the media.'' Examples of events would include shooting down of the Malaysia Airlines plane over Ukraine on July 18th, 2014 and HSBC's admittance of aiding their clients in tax evasion on February 9th, 2015 (Figure~\ref{fig:event2}). Events such as these are covered by many articles and the question is how to find all the articles in different languages that are describing a single event. Generally, events are more specific than general themes as the time component plays an important role -- for example, the two wars in Iraq would be considered as separate events.

As input, we consider a stream of articles in different languages and a list of events. Our goal is to assign articles to their corresponding events. A priori, we do not know the coverage of the articles, that is, not all the events may be covered and we do not know that all the articles necessarily fit into one of the events. The task is divided into two parts: detecting events within each language and then linking events across languages. In this paper we address the second step.

 We consider a high volume of articles in different languages. By using a language detector, the stream is split into separate monolingual streams. Within each monolingual stream, an online clustering approach is employed, where tracked clusters correspond to our definition of events - this is based on the Event Registry system~\cite{Leban2014W,Leban2014I}. Our main goal in this paper is to connect such clusters (representations of events) across languages, that is, to detect that a set of articles in language $A$ reports on the same event as a set of articles in language $B$.

Our approach to link clusters across languages combines two ingredients: a cross-lingual document similarity measure, which can be interpreted as a language independent topic model, and semantic annotation of documents, which enables an alternative way to comparing documents. Since this work represents a complicated pipeline, we concentrate on these two specific elements. Overall, the approach should be considered from a systems' perspective (considering the system as a whole) rather than considering these problems in isolation.

The first ingredient of our approach to link clusters across languages represents a continuation of previous work~\cite{nips,sikdd,nips2,iti} where we explored representations of documents which were valid over multiple languages.  The representations could be interpreted as multilingual topics, which were then used as proxies to compute cross-lingual similarities between documents. To learn the representations, we use Wikipedia as a training corpus. Significantly, we do not only consider the major or \emph{hub} languages such as English, German, French, etc. which have significant overlap in article coverage, but also smaller languages (in terms of number of Wikipedia articles) such as Slovenian and Hindi, which may have a negligible overlap in article coverage. We can then define a similarity between any two articles regardless of language,  which allows us to cluster the articles according to topic. The underlying assumption is that articles describing the same event are similar  and will therefore be put into the same cluster.

Based on the similarity function, we propose a novel algorithm for linking events/clusters across languages. The approach is based on learning a classification model from labelled data based on several sets of features. In addition to these features, cross-lingual similarity is also used to quickly identify a small list of potential linking candidates for each cluster.
This greatly increases the scalability of the system.

The paper is organized as follows: we first provide an overview of the system as a whole in Section~\ref{sec:pipeline}, which includes a subsection that summarizes the main system requirements. We then present related work in Section~\ref{sec:related}. The related work covers work on cross-lingual document similarity as well as work on cross-lingual cluster linking. In Section~\ref{sec:crosslingual}, we introduce the problem of cross-lingual document similarity computation and describe several approaches to the problem, most notably a new approach based on hub languages.
%The approach in Section~\ref{sec:kmeans} serves as a baseline, and the approaches in Section~\ref{sec:CCA} and Section~\ref{sec:LSI} serve as the basis on which we proposed a new approach presented in Section~\ref{sec:hublang}
 In Section~\ref{sec:linking}, we introduce the central problem of cross-lingual linking of clusters of news articles and our approach that combines the cross-lingual similarity functions with knowledge extraction based techniques. Finally, we present and interpret the experimental results in Section~\ref{sec:evaluation} and discuss conclusions and point out several promising future directions.

\section{Pipeline}\label{sec:pipeline}

\begin{figure}[tbp]
\centering
\includegraphics[width=\textwidth]{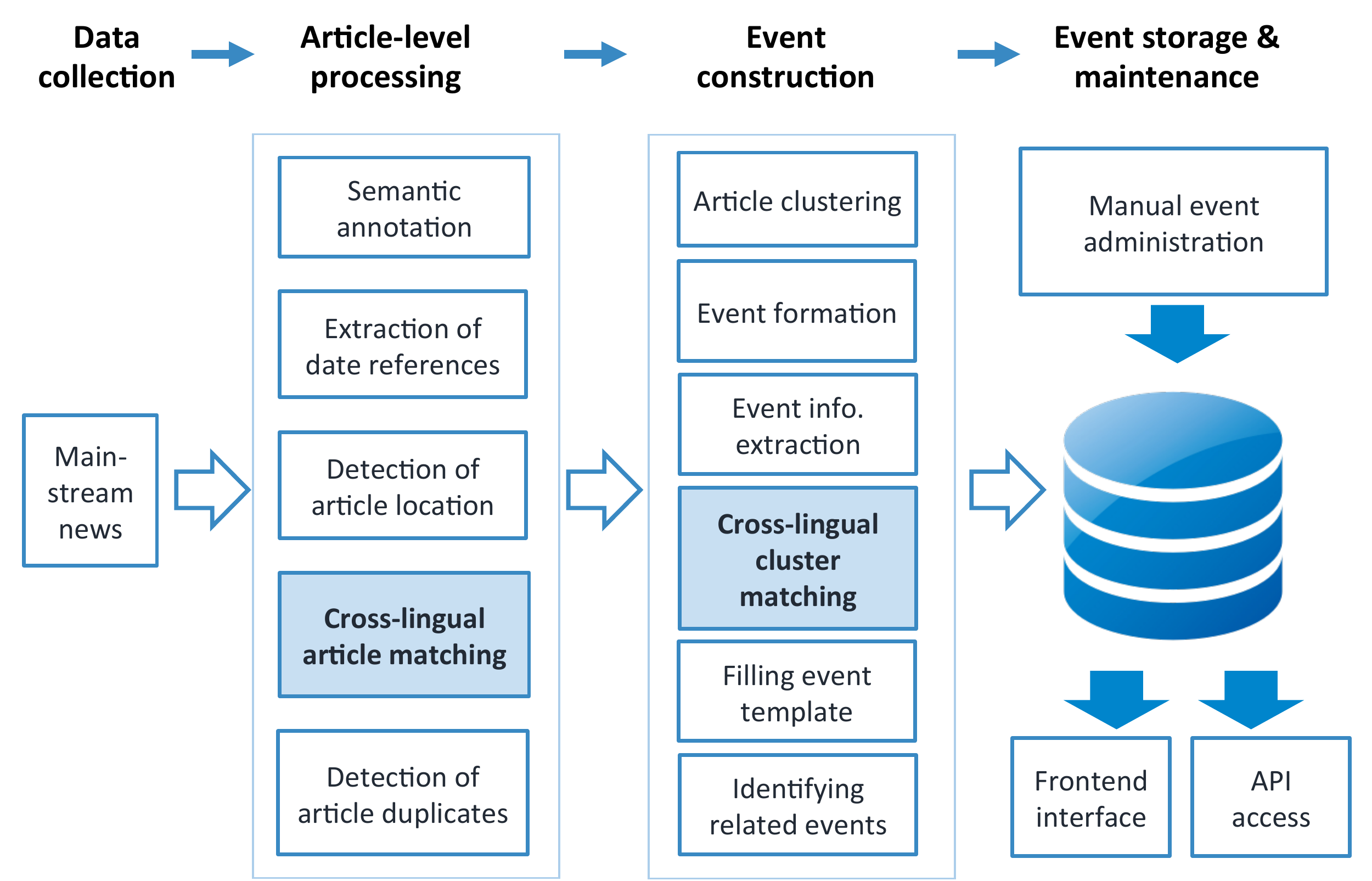}
\caption{\label{fig:erpipeline}  The Event Registry pipeline. After new articles are collected, they are first analyzed individually (Article-level processing). In the next step, groups of articles about the same event are identified and relevant information about the event is extracted (Event construction phase). Although the pipeline contains several components, we focus only on the two highlighted in the image.}
\end{figure}

We base our techniques of cross-lingual event linking on an online system for detection of world events, called Event Registry~\cite{Leban2014W,Leban2014I}. Event Registry is a repository of events, where events are automatically identified by analyzing news articles that are collected from numerous news outlets all over the world. The important components in the pipeline of the Event Registry are shown in Figure~\ref{fig:erpipeline}. We will now briefly describe the main components.

The collection of the news articles is performed using the Newsfeed service~\cite{Trampus2012}. The service monitors RSS feeds of around 100,000 mainstream news outlets available globally. Whenever a new article is detected in the RSS feed, the service downloads all available information about the article and sends the article through the pipeline. Newsfeed downloads daily on average around 200,000 news articles in various languages, where English, Spanish and German are the most common.

Collected articles are first semantically annotated by identifying mentions of relevant concepts -- either entities or important keywords. The disambiguation and entity linking of the concepts is done using Wikipedia as the main knowledge base. The algorithm for semantic annotation uses machine learning to detect significant terms within unstructured text and link them to the appropriate Wikipedia articles. The approach models link probability and combines prior word sense distributions with context based sense distributions. The details are reported in~\cite{Milne:2008:LLW:1458082.1458150} and ~\cite{zhang2014saaacamactat}. As a part of the semantic annotation we also analyze the dateline of the article to identify the location of the described event as well as to extract dates mentioned in the article using a set of regular expressions. Since articles are frequently revised we also detect if the collected article is just a revision of a previous one so that we can use this information in next phases of the pipeline. The last important processing step on the document level is to efficiently identify which articles in other available languages are most similar to this article. The methodology for this task is one of the main contributions of this paper and is explained in details in Section~\ref{sec:crosslingual}.

As the next step, an online clustering algorithm is applied to the articles in order to identify groups of articles that are discussing the same event. For each new article, the clustering algorithm determines if the article should be assigned to some existing cluster or into a new cluster. The underlying assumption is that articles that are describing the same event are similar enough and will therefore be put into the same cluster. For clustering, each new article is first tokenized, stop words are removed and the remaining words are stemmed.
The remaining tokens are represented in a vector-space model and normalized using TF-IDF\footnote{The IDF weights are dynamically computed for each new article over all news articles within a 10 day window.} (see Section~\ref{sec:tfidf} for the definition). Cosine similarity is used to find the most similar existing cluster, by comparing the document's vector to the centroid vector of each cluster. A user-defined threshold is used to determine if the article is not similar enough to any existing clusters (0.4 was used in our experiments).  If the highest similarity is above the threshold, the article is assigned to the corresponding cluster, otherwise a new cluster is created, initially containing only the single article. Whenever an article is assigned to a cluster, the cluster's centroid vector is also updated. Since articles about an event are commonly written only for a short period of time, we remove clusters once the oldest article in the cluster becomes more than 4 days old. This housekeeping mechanism prevents the clustering from becoming slow and also ensures that articles are not assigned to obsolete clusters. Details of the clustering approach are described in~\cite{brank2014}.

Once the number of articles in a cluster reaches a threshold (which is a language dependent parameter), we assume that the articles in the cluster are describing an event. At that point, a new event with a unique ID is created in Event Registry, and the cluster with the articles is assigned to it. By analyzing the articles, we extract the main information about the event, such as the event location, date, most relevant entities and keywords, etc.

Since articles in a cluster are in a single language, we also want to identify any other existing clusters that report about the same event in other languages and join these clusters into the same event. This task is performed using a classification approach which is the second major contribution of this paper. It is described in detail in Section~\ref{sec:linking}.

When a cluster is identified and information about the event is extracted, all available data is stored in a custom-built database system. The data is then accessible through the API or a web interface at \protect\url{http://eventregistry.org/}, which provide numerous search and visualization options.

\subsection{System Requirements}
\label{sec:sysreq}

Our goal is to build a system that monitors global media and analyzes how events are being reported on. Our approach consists of two steps: tracking events separately in each language (based on language detection and an online clustering approach) and then connecting them. The pipeline must be able to process millions of articles per day and perform billions of similarity computations each day. Both steps rely heavily on similarity computation, which implies that this must be highly scalable.

Therefore, we focus on implementations that run on a single shared memory machine, as opposed to clusters of machines. This simplifies implementation and system maintenance.
To summarize, the following properties are desirable:
\begin{itemize}
\item \textbf{Training} - The training (building cross-lingual models) should scale to many languages and should be robust to the quality of training resources. The system should be able to take advantage of comparable corpora (as opposed to parallel translation-based corpora), with missing data.
\item \textbf{Operation efficiency} - The similarity computation should be fast - the system must be able to handle billions of similarity computations per day. Computing the similarity between a new document and a set of known documents should be efficient (the main application is linking documents between two monolingual streams).
\item \textbf{Operation cost} - The system should run on a strong shared machine server and not rely on paid services.
\item \textbf{Implementation} - The system is straightforward to implement, with few parameters to tune.
\end{itemize}
We believe that a cross-lingual similarity component that meets such requirements is very desirable in a commercial setting, where several different costs have to be taken into consideration.

\section{Related work}\label{sec:related}

In this section, we describe previous work described  in the literature. Since there are two distinctive tasks that we tackle in this paper (computing cross-lingual document similarity and cross-lingual cluster linking), we have separated the related work into two corresponding parts.

\subsection{Cross-lingual document similarity}

There are four main families of approaches to cross-lingual similarity.

\textbf{Translation and dictionary based}. The most obvious way to compare documents written in different languages is to use machine translation and perform monolingual similarity, see  ~\cite{multilingualBook,plagiarism} for several variations of translation based approaches. One can use free tools such as Moses~\cite{moses} or translation services, such as Google Translate (\protect\url{https://translate.google.com/}). There are two issues with such approaches: they solve a harder problem than needs to be solved and they are less robust to training resource quality - large sets of translated sentences are typically needed. Training Moses for languages with scarce linguistic resources is thus problematic. The issue with using online services such as Google Translate is that the APIs are limited and not free. The operation efficiency and cost requirements make translation-based approaches less suited for our system. Closely related are works Cross-Lingual Vector Space Model (CL-VSM)~\cite{plagiarism} and the approach presented in~\cite{pouliquen2008story} which both compare documents by using dictionaries, which in both cases are EuroVoc dictionaries~\cite{eurovoc}. The generality of such approaches is limited by the quality of available linguistic resources, which may be scarce or non-existent for certain language pairs.

\textbf{Probabilistic topic models}. There exist many variants to modelling documents in a language independent way by using probabilistic graphical models. The models include:  Joint Probabilistic Latent Semantic Analysis (JPLSA)~\cite{platt2010translingual}, Coupled Probabilistic LSA (CPLSA)~\cite{platt2010translingual}, Probabilistic Cross-Lingual LSA (PCLLSA)~\cite{PCL_LSA} and Polylingual Topic Models (PLTM)~\cite{polyLDA} which is a Bayesian version of PCLLSA. The methods (except for CPLSA) describe the multilingual document collections as samples from generative probabilistic models, with variations on the assumptions on the model structure. The topics represent latent variables that are used to generate observed variables (words), a process specific to each language. The parameter estimation is posed as an inference problem which is typically intractable and one usually solves it using approximate techniques. Most variants of solutions are based on Gibbs sampling or Variational Inference, which are nontrivial to implement and may require an experienced practitioner to be applied. Furthermore, representing a new document as a mixture of topics is another potentially hard inference problem which must be solved.

\textbf{Matrix factorization}. Several matrix factorization based approaches exist in the literature. The models include: Non-negative matrix factorization based~\cite{nonnegfactor_lsi}, Cross-Lingual Latent Semantic Indexing CL-LSI~\cite{cl_lsi,multilingualBook}, Canonical Correlation Analysis (CCA)~\cite{Hotelling}, Oriented Principal Component Analysis (OPCA)~\cite{platt2010translingual}. The quadratic time and space dependency of the OPCA method makes it impractical for large scale purposes. In addition, OPCA forces the vocabulary sizes for all languages to be the same, which is less intuitive. For our setting, the method in ~\cite{nonnegfactor_lsi} has a prohibitively high computational cost when building models (it uses dense matrices whose dimensions are a product of the training set size and the vocabulary size). Our proposed approach combines CCA and CL-LSI. Another closely related method is Cross-Lingual Explicit Semantic Analysis (CL-ESA)~\cite{ESA}, which uses Wikipedia (as do we in the current work) to compare documents. It can be interpreted as using the sample covariance matrix between features of two languages to define the dot product which is used to compute similarities.
The authors of CL-ESA compare it to CL-LSI and find that CL-LSI can outperform CL-ESA in an information retrieval, but is costlier to optimize over a large corpus (CL-ESA requires no training). We find that the scalability argument does not apply in our case: based on advances in numerical linear algebra we can solve large CL-LSI problems (millions of documents as opposed to the 10,000 document limit reported in~\cite{ESA}). In addition, CL-ESA is less suited for computing similarities between two large monolingual streams. For example, each day we have to compute similarities between 500,000 English and 500,000 German news articles. Comparing each German news article with 500,000 English news articles is either prohibitively slow (involves projecting all English articles on Wikipedia) or consumes too much memory (involves storing the projected English articles, which for a Wikipedia of size 1,000,000 is a 500,000 by 1000,0000 non-sparse matrix).

\textbf{Monolingual}. Finally, related work includes monolingual approaches that treat document written in different languages in a monolingual fashion. The intuition is that named entities (for example, ``Obama'') and cognate words (for example, ``tsunami'') are written in the same or similar fashion in many languages. For example, the Cross-Language Character n-Gram Model (CL-CNG)~\cite{plagiarism} represents documents as bags of character $n$-grams. Another approach is to use language dependent keyword lists based on cognate words ~\cite{pouliquen2008story}. These approaches may be suitable for comparing documents written in languages that share a writing system, which does not apply to the case of global news tracking.

Based on our requirements in Section~\ref{sec:sysreq}, we chose to focus on methods based on vector space models and linear embeddings. We propose a method that is more efficient than popular alternatives (a clustering-based approach and latent semantic indexing), but is still simple to optimize and use.

\subsection{Cross-lingual cluster linking}

Although there are a number of services that aggregate news by identifying clusters of similar articles, there are almost no services that provide linking of clusters over different languages. Google News as well as Yahoo! News are able to identify clusters of articles about the same event, but offer no linking of clusters across languages. The only service that we found, which provides cross-lingual cluster linking, is the European Media Monitor (EMM)~\cite{pouliquen2008story,Steinberger2008}. EMM clusters articles in 60 languages and then tries to determine which clusters of articles in different languages describe the same event. To achieve cluster linking, EMM uses three different language independent vector representations for each cluster. The first vector contains the weighted list of references to countries mentioned in the articles, while the second vector contains the weighted list of mentioned people and organizations. The last vector contains the weighted list of Eurovoc subject domain descriptors. These descriptors are topics, such as \emph{air transport}, \emph{EC agreement}, \emph{competition} and \emph{pollution control} into which articles are automatically categorized~\cite{Pouliquen2006}. Similarity between clusters is then computed using a linear combination of the cosine similarities computed on the three vectors. If the similarity is above the threshold, the clusters are linked. Compared to EMM, our approach uses document similarities to obtain a small set of potentially equivalent clusters. Additionally, we do not decide if two clusters are equivalent based on a hand-set threshold on a similarity value -- instead we use a classification model that uses a larger set of features related to the tested pair of clusters.

A system, which is significantly different but worth mentioning, is the GDELT project~\cite{Leetaru2013Gdelt}. In GDELT, events are also extracted from articles, but in their case, an event is specified in a form of a triple containing two actors and a relation. The project contains an extensive vocabulary of possible relations, mostly related to political events. In order to identify events, GDELT collects articles in more than 65 languages and uses machine translation to translate them to English. All information extraction is then done on the translated article.

\section{Cross-lingual Document Similarity}
\label{sec:crosslingual}
Document similarity is an important component in techniques from text mining and natural language processing. Many techniques use the similarity as a black box, e.g., a kernel in Support Vector Machines. Comparison of documents (or other types of text snippets) in a monolingual setting is a well-studied problem in the field of information retrieval ~\cite{Salton88term-weightingapproaches}. We first formally introduce the problem followed by a description of  our approach.

\subsection{Problem definition}\label{sec:tfidf}
We will first describe how documents are represented as vectors and how to compare documents in a mono-lingual setting. We then define a way to measure cross-lingual similarity which is natural for the models we consider.

\noindent\textbf{Document representation.}
The standard vector space model~\cite{Salton88term-weightingapproaches} represents documents as vectors, where each term corresponds to a word or a phrase in a fixed vocabulary. Formally, document $d$ is represented by a vector $x \in \RR^n$, where $n$ corresponds to the size of the vocabulary, and vector elements $x_k$ correspond to the number of times term $k$ occurred in the document, also called \emph{term frequency} or $TF_k(d)$.

We also used a term re-weighting scheme that adjusts for the fact that some words occur more frequently in general. A term weight should correspond to the importance of the term for the given corpus. The common weighting scheme is called \emph{Term Frequency Inverse Document Frequency} ($TFIDF$) weighting. An \emph{Inverse Document Frequency} ($IDF$) weight for the dictionary term $k$ is defined as $\log\left( \frac{N}{DF_k} \right)$, where $DF_k$ is the number of documents in the corpus which contain term $k$.
When building cross-lingual models, the IDF scores were computed with respect to the Wikipedia corpus. In the other part of our system, we computed TFIDF vectors on streams of news articles in multiple languages. There the IDF scores for each language changed dynamically - for each new document we computed the IDF of all news articles within a 10 day window.

Therefore we can define a document's $TFIDF$ as
$$ x_{ij}  := \frac{\mbox{term frequency in document } i}{\mbox{inverse document frequency of term } j}.$$
The $TFIDF$ weighted vector space model document representation corresponds to a map $\phi : \text{text} \rightarrow \RR^n$ defined by:
$$\phi(d)_k = {TF}_k(d) \log\left( \frac{N}{{DF}_k}\right).$$

\noindent\textbf {Mono-lingual similarity.}
A common way of computing similarity between documents is \emph{cosine similarity},
$$sim(d_1, d_2) = \frac{\langle \phi(d_1), \phi(d_2)\rangle}{\|\phi(d_1)\| \|\phi(d_2)\|},$$
where $\langle \cdot,\cdot \rangle$ and $\|\cdot\|$ are standard inner product and Euclidean norm. When dealing with two or more languages, one could ignore the language information
and build a vector space using the union of tokens over the languages. A cosine similarity function in such a space can be useful to some extent, for example ``Internet'' or ``Obama'' may appear both in Spanish and English texts and the presence of such terms in both an English and a Spanish document would contribute to their similarity. In general however, large parts of vocabularies may not intersect. This means that given a language pair, many words in both languages cannot contribute to the similarity score. Such cases can make the similarity function very insensitive to the data.

\noindent\textbf {Cross-lingual similarity.}
Processing a multilingual dataset results in several vector spaces with varying dimensionality, one for each language. The dimensionality of the vector space corresponding to the $i$-th language is denoted by $n_i$ and the vector space model mapping is denoted by $\phi_i : \text{text} \rightarrow \RR^{n_i}$.
The similarity between documents in language $i$ and language $j$ is defined as a bilinear operator represented as a matrix $S_{i,j} \in \RR^{n_i \times n_j}$:
$$sim_{i,j}(d_1, d_2) = \frac{ \langle \phi_i (d_1), S_{i,j} \phi_j (d_2) \rangle }{\|\phi_i(d_1)\| \|\phi_j(d_2)\|},$$
where $d_1$ and $d_2$ are documents written in the $i$-th and $j$-th language respectively. If the maximal singular value of $S_{i,j}$ is bounded by $1$, then the similarity scores will lie on the interval $[-1, 1]$. We will provide an overview of the models in Section \ref{sec:models} and then introduce additional notation in \ref{sec:notation}. Starting with Section \ref{sec:kmeans} and ending with Section \ref{sec:hublang} we will describe some approaches to compute $S_{i,j}$ given training data.

\subsection{Cross-Lingual Models}\label{sec:models}
In this section, we will describe several approaches to the problem of computing the multilingual similarities introduced in Section~\ref{sec:tfidf}. We present four approaches:
a simple approach based on $k$-means clustering in Section~\ref{sec:kmeans}, a standard approach based on singular value decomposition in Section~\ref{sec:LSI}, a related
approach called Canonical Correlation Analysis (CCA) in Section~\ref{sec:CCA} and finally a new method, which is an extension of CCA to more than two languages in Section~\ref{sec:hublang}.
CCA can be used to find correlated patterns for a pair of languages, whereas the extended method optimizes a
Sum of Squared Correlations (SSCOR) between several language pairs, which was introduced in~\cite{Kettenring}. The SSCOR problem is difficult to solve in our setting (hundreds of thousands of features, hundreds of thousands of examples). To tackle this, we propose a method which consists of two ingredients.
 The first one is based on an observation that certain datasets (such as Wikipedia) are biased towards one language (English for Wikipedia), which can be exploited
 to reformulate a difficult optimization problem as an eigenvector problem. The second ingredient is dimensionality reduction using CL-LSI, which
 makes the eigenvector problem computationally and numerically tractable.

We concentrate on approaches that are based on linear maps rather than alternatives, such as machine translation and probabilistic models, as discussed in the section on related work.
We will start by introducing some notation.

\subsection{Notation}\label{sec:notation}

The cross-lingual similarity models presented in this paper are based on comparable corpora. A \emph{comparable corpus} is a collection of documents in multiple languages, with alignment between documents that are of the same topic, or even a rough translation of each other. Wikipedia is an example of a comparable corpus, where a specific entry can be described in multiple languages (e.g., ``Berlin" is currently described in 222 languages). News articles represent another example, where the same event can be described by newspapers in several languages.

\begin{figure}[tbp]
\centering
\includegraphics[width=9cm]{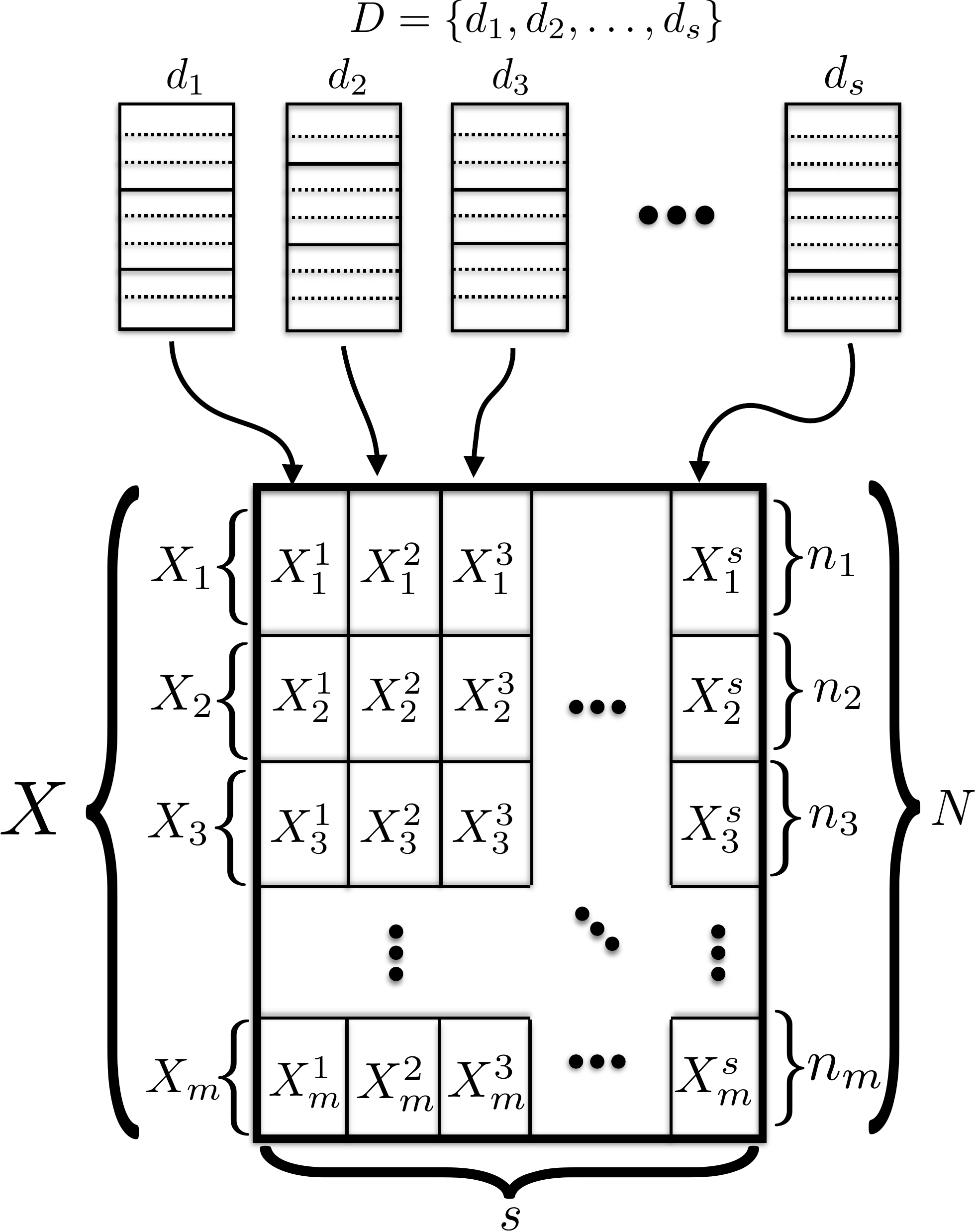}
\caption{\label{fig:stacked_matrices} Multilingual corpora and their matrix representations using the vector space model.}
\end{figure}

More formally, a \emph{multilingual document} $d = (u_1,\ldots u_m)$ is a tuple of $m$ documents on the same topic (comparable), where $u_i$ is the document written in language $i$. Note that an individual document $u_i$ can be an empty document (missing resource) and each $d$ must contain at least \textbf{two nonempty documents}. This means that in our analysis we discard strictly monolingual documents for which no cross-lingual information is available. A comparable corpus $D = {d_1, \ldots, d_s}$ is a collection of $s$ multilingual documents. By using the vector space model, we can represent $D$ as a set of $m$ matrices $X_1,\ldots,X_m$, where $X_i \in \RR^{n_i \times s}$ is the matrix corresponding to the language $i$ and $n_i$ is the vocabulary size of language $i$. Furthermore, let $X_i^{\ell}$ denote the $\ell$-th column of matrix $X_i$ and the matrices respect the document alignment - the vector $X_i^\ell$ corresponds to the TFIDF vector of the $i$-th component of multilingual document $d_\ell$. We use $N$ to denote the total row dimension of $X$, i.e., $N:= \sum_{i=1}^m n_i$. See Figure~\ref{fig:stacked_matrices} for an illustration of the introduced notation.

We will now describe four models to cross-lingual similarity computation in the next sub-sections.
\subsection{$k$-means}\label{sec:kmeans}

\begin{figure}[tbp]
\centering
\includegraphics[width=10cm]{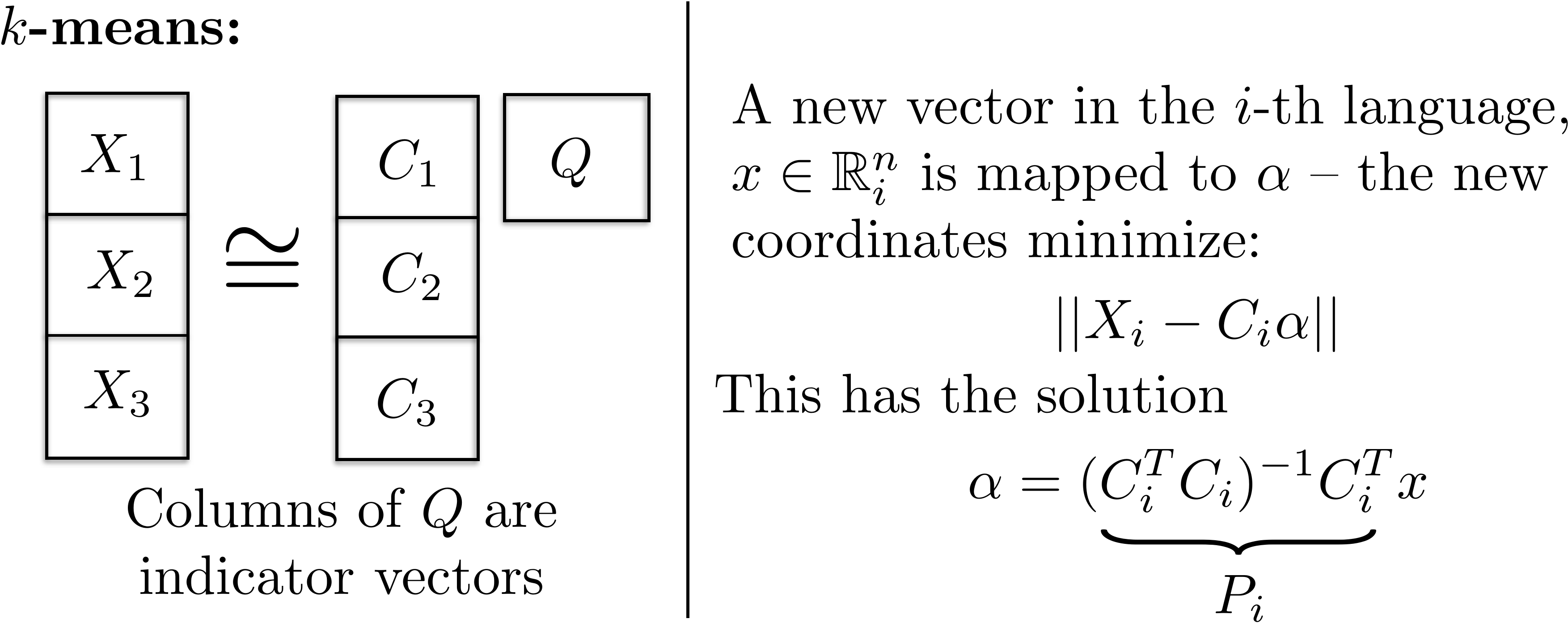}
\caption{\label{fig:kmeans} $k$-means algorithm and coordinate change.}
\end{figure}

The $k$-means algorithm is perhaps the most well-known and widely-used clustering algorithm. Here, we present its application
to compute cross-lingual similarities. The idea is based on concatenating the corpus matrices, running standard $k$-means clustering to obtain the matrix of centroids, ``reversing" the concatenation step to obtain a set of aligned bases, which are finally used to compute cross-lingual similarities. See Figure~\ref{fig:kmeans} for overview of the procedure. The left side of Figure~\ref{fig:kmeans} illustrates the decomposition and the right side summarizes the coordinate change.

 In order to apply the algorithm, we first merge all the term-document matrices into a single matrix $X$ by stacking the individual term-document matrices (as seen in Figure~\ref{fig:stacked_matrices}):
$$X := \begin{bmatrix}X_1^T ,X_2^T, \cdots, X_m^T \end{bmatrix}^T,$$
such that the columns respect the alignment of the documents (here MATLAB notation for concatenating matrices is used). Therefore, each document  is represented by a long vector indexed by the terms in all languages.

We then run the $k$-means algorithm~\cite{kmeans} and obtain a centroid matrix $C \in \RR^{N \times k}$, where the $k$ columns represent centroid vectors. The centroid matrix can be split vertically into $m$ blocks: $$C = [C_1^T \cdots C_m^T]^T,$$ according to the number of dimensions of each language, i.e., $C_i \in \RR^{n_i \times k}$.
To reiterate, the matrices $C_i$ are computed using a multilingual corpus matrix $X$ (based on Wikipedia for example).

To compute  cross-lingual document similarities on new documents, note that each matrix $C_i$ represents a vector space basis and can be used to map points in $\RR^{n_i}$ into a $k$-dimensional space, where the new coordinates of a vector $x \in \RR^{n_i}$ are expressed as: $$(C_i^T C_i)^{-1} C_i^T x_i.$$

The resulting matrix for similarity computation between language $i$ and language $j$ is defined up to a scaling factor as:
$$C_i(C_i^T C_i)^{-1} (C_j^T C_j)^{-1} C_j.$$

The matrix is a result of mapping documents in a language independent space using pseudo-inverses of the centroid matrices $P_i = (C_i^T C_i)^{-1} C_i$ and then comparing them using the standard inner product, which results in the matrix $P_i^T P_j$. For the sake of presentation, we assumed that the centroid vectors are linearly independent. (An independent subspace could be obtained using an additional Gram-Schmidt step~\cite{golub} on the matrix $C$, if this was not the case.)

\subsection{Cross-Lingual Latent Semantic Indexing}\label{sec:LSI}

The second approach we consider is Cross-Lingual Latent Semantic Indexing (CL-LSI)~\cite{cl_lsi} which is a variant of LSI~\cite{lsi} for more than one language. The approach is very similar to $k$-means, where we first concatenate the corpus matrices, compute a decomposition, which in case of CL-LSI is a truncated Singular Value Decomposition (SVD), decouple the
 column space matrix and use the blocks to compute linear maps to a common vector space, where standard cosine similarity is used to compare documents.

\begin{figure}[tbp]
\centering
\includegraphics[width=10cm]{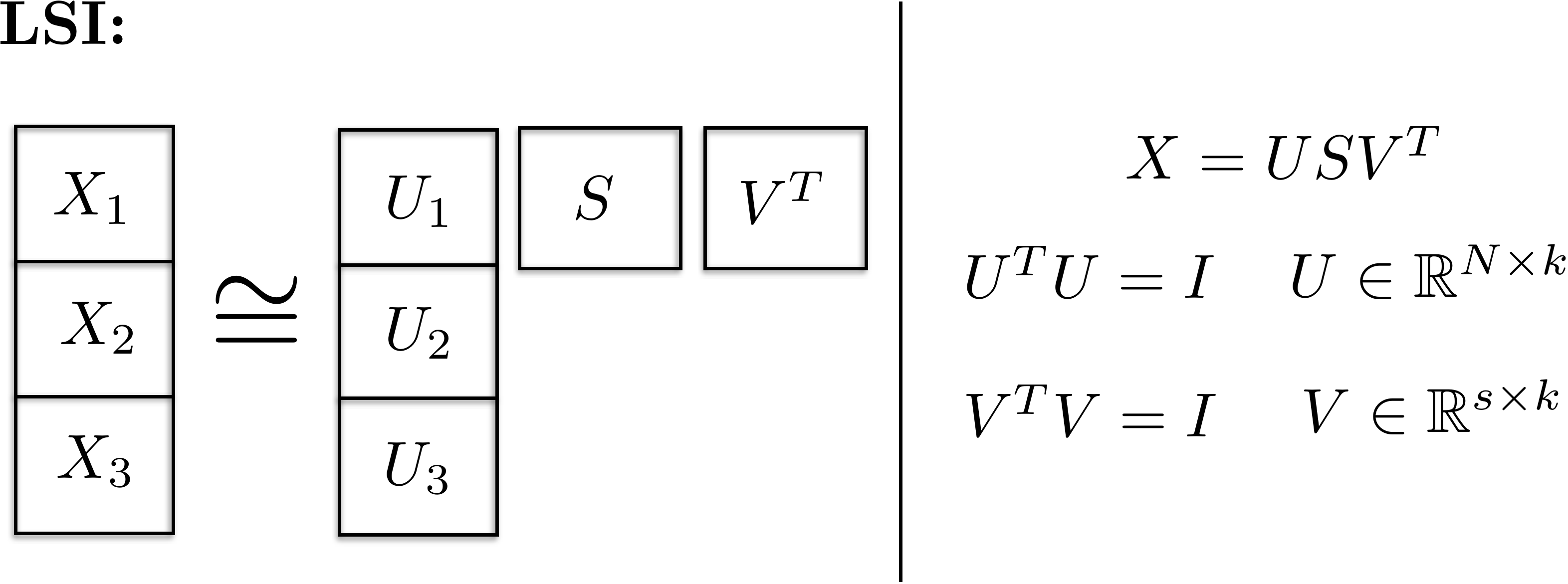}
\caption{\label{fig:lsi} LSI multilingual corpus matrix decomposition.}
\end{figure}

 The method is based on computing a truncated singular value decomposition of the concatenated corpus matrix $X \approx U S V^T$. See Figure~\ref{fig:lsi} for the decomposition. Representing documents in ``topic`` coordinates is done in the same way as in the $k$-means case (see Figure~\ref{fig:kmeans}), we will describe how to compute the coordinate change functions.

The cross-lingual similarity functions are based on a rank-$k$ truncated SVD: $X \approx U \Sigma V^T,$ where $U \in \RR^{N \times k}$ are basis vectors of interest and $\Sigma \in \RR^{k \times k}$ is a truncated diagonal matrix of singular eigenvalues. An aligned basis is obtained by first splitting $U$ vertically according to the number of dimensions of each language: $U = [U_1^T \cdots U_m^T]^T$. Then, the same as with $k$-means clustering, we compute the pseudoinverses $P_i = (U_i^T U_i)^{-1} U_i^T$. The matrices $P_i$ are used to change the basis from the standard basis in $\RR^{n_i}$ to the basis spanned by the columns of $U_i$.

\paragraph{Implementation note}

  Since the matrix $X$ can be large we could use an iterative method like the Lanczos algorithm with reorthogonalization~\cite{golub} to find the left singular vectors (columns of $U$) corresponding to the largest singular values. It turns out that the Lanczos method converges slowly as the gap between the leading singular values is small. Moreover, the Lanczos method is hard to parallelize. Instead, we use a randomized version of the SVD~\cite{tropp} that can be viewed as a block Lanczos method. That enables us to use parallelization and speeds up the computation considerably.

To compute the matrices $P_i$ we used the QR algorithm~\cite{golub} to factorize $U_i$ as $U_i = Q_i R_i$, where $Q_i^TQ_i = I$ and $R_i$ is a triangular matrix. $P_i$ is then obtained by solving $R_i P_i = Q_i$.

\subsection{Canonical Correlation Analysis}\label{sec:CCA}
 We now present a statistical technique to analyze data from two sources, an extension of which will be presented in the next section.
 Canonical Correlation Analysis (CCA)~\cite{Hotelling} is a dimensionality reduction technique similar to Principal Component Analysis (PCA)~\cite{Pearson1901On}, with the additional assumption that the data consists of feature vectors that arose from two sources (two views) that share some information. Examples include: bilingual document collection ~\cite{mrpqr} and collection of images and captions~\cite{Hardoon_usingimage}. Instead of looking for linear combinations of features that maximize the variance (PCA) we look for a linear combination of feature vectors from the first view and a linear combination for the second view, that are maximally correlated.

Interpreting the columns of $X_i$ as observation vectors sampled from an underlying distribution $\mathcal{X}_i \in \RR^{n_i}$, the idea is to find two weight vectors $w_i \in \RR^{n_i}$ and $w_j \in \RR^{n_j}$ so that the random variables $w_i^T \cdot \mathcal{X}_i$ and $w_j^T \cdot \mathcal{X}_j$ are maximally correlated ($w_i$ and $w_j$ are used to map the random vectors to random variables, by computing weighted sums of vector components). Let $\rho(x,y)$ denote the sample-based correlation coefficient between two vectors of observations $x$ and $y$. By using the sample matrix notation $X_i$ and $X_j$ (assuming no data is missing to simplify the exposition), this problem can be formulated as the following optimization problem:
\begin{equation*}
\begin{aligned}
& \underset{w_i \in \RR^{n_i}, w_j \in \RR^{n_j}}{\text{maximize}}
& & \rho(w_i^T X_i , w_j^T X_j) = \frac{w_i^T C_{i,j} w_j}{\sqrt{w_i^T C_{i,i} w_i} \sqrt{w_j^T C_{j,j} w_j}},
\end{aligned}
\end{equation*}
where $C_{i,i}$ and $C_{j,j}$ are empirical estimates of variances of $\mathcal{X}_i$ and $\mathcal{X}_j$ respectively and $C_{i,j}$ is an estimate for the covariance matrix. Assuming that the observation vectors are centered (only for the purposes of presentation), the matrices are computed in the following way: $C_{i,j} = \frac{1}{n-1}X_i X_j^T$, and similarly for $C_{i,i}$ and $C_{j,j}$.
The optimization problem can be reduced to an eigenvalue problem and includes inverting the variance matrices $C_{i,i}$ and $C_{j,j}$. If the matrices are not invertible, one can use a regularization technique by replacing $C_{i,i}$ with $(1- \kappa)C_{i,i} + \kappa I$, where $\kappa \in [0,1]$ is the regularization coefficient and $I$ is the identity matrix. (The same can be applied to $C_{j,j}$.)
A single canonical variable is usually inadequate in representing the original random vector and typically one looks for $k$ projection pairs $(w_i^1, w_j^1),\ldots,(w_i^k, w_j^k)$, so that $(w_i^{u})^T \mathcal{X}_i$ and $(w_j^{u})^T \mathcal{X}_j$ are highly correlated and $(w_i^{u})^T \mathcal{X}_i$ is uncorrelated with $(w_i^{v})^T \mathcal{X}_i$  for $u \neq v$ and analogously for $w_j^u$ vectors.

Note that the method in its original form is only applicable to two languages where an aligned set of observations is available. The next section will describe a scalable extension of CCA to more than two languages.

\subsection{Hub language based CCA Extension}\label{sec:hublang}
Building cross-lingual similarity models based on comparable corpora is challenging for two main reasons. The first problem is related to missing alignment data: when a number of languages is large, the dataset of documents that cover all languages is small (or may even be empty). Even if only two languages are considered, the set of aligned documents can be small (an extreme example is given by the Piedmontese and Hindi Wikipedias where no inter-language links are available), in which case none of the methods presented so far are applicable.
 The second challenge is scale - the data is high-dimensional (many languages with hundreds of thousands of features per language) and the number of multilingual documents may be large (over one million in case of Wikipedia). The optimization problem posed by CCA is not
 trivial to solve: the covariance matrices themselves are prohibitively large to fit in memory (even storing a 100,000 by 100,000 element matrix requires 80GB of memory) and iterative matrix-multiplication based approaches to  solving generalized eigenvalue problems are required (the covariance matrices can be expressed as products of sparse matrices, which means we have fast matrix-vector multiplication).

We now describe an extension of CCA to more than two languages, which can be trained on large comparable corpora and can handle missing data.
 The extension we consider is based on a generalization of CCA to more than two views, introduced in~\cite{Kettenring}, namely the Sum of Squared Correlations SSCOR, which we will state formally later in this section. Our approach exploits a certain characteristic of the data, namely the \emph{hub language} characteristic (see below) in two
 ways: to reduce the dimensionality of the data and to simplify the optimization problem.

\paragraph{Hub language characteristic.}
In the case of Wikipedia, we observed that even though the training resources are scarce for certain language pairs, there often exists indirect training data. By considering a third language, which has training data with both languages in the pair,  we can use the composition of learned maps as a proxy. We refer to this third language as a hub language.

A \emph{hub language} is a language with a high proportion of non-empty documents in $D = \left\{d_1,..., d_{\ell}\right\}$. As we have mentioned, we only focus on multilingual documents that include at least two languages. The prototypical example in the case of Wikipedia is English. Our notion of the hub language could be interpreted in the following way.
If a non-English Wikipedia page contains one or more links to variants of the page in other languages, English is very likely to be one of them. That makes English a hub language.

We use the following notation to define subsets of the multilingual comparable corpus: let $a(i,j)$ denote the index set of all multilingual documents with non-missing data for the $i$-th and $j$-th language:  $$a(i,j) = \left\{k~ |~ d_k = (u_1,...,u_m), u_i \neq \emptyset, u_j \neq \emptyset \right\},$$ and let $a(i)$ denote the index set of all multilingual documents with non missing data for the $i$-th language.

We now describe a two step approach to building a cross-lingual similarity matrix. The first part is related to LSI and reduces the dimensionality of the data. The second step refines the linear mappings and optimizes the linear dependence between data.

\paragraph{Step 1: Hub language based dimensionality reduction.}

The first step in our method is to project $X_1, \ldots, X_m$ to lower-dimensional spaces without destroying the cross-lingual structure. Treating the nonzero columns of $X_i$ as observation vectors sampled from an underlying distribution $\mathcal{X}_i \in V_i = \RR^{n_i}$, we can analyze the empirical cross-covariance matrices:
$$C_{i,j} = \frac{1}{|a(i,j)|-1 }\sum_{\ell \in a(i,j)} (X_i^{\ell} - c_i)\cdot (X_j^{\ell} - c_j)^T,$$
 where $c_i = \frac{1}{a_i} \sum_{\ell \in a(i)}X_i^{\ell}$. By finding low-rank approximations of $C_{i,j}$ we can identify the subspaces of $V_i$ and $V_j$ that are relevant for extracting linear patterns between $\mathcal{X}_i$ and $\mathcal{X}_j$. Let $X_1$ represent the hub language corpus matrix. The LSI approach to finding the subspaces is to perform the singular value decomposition on the full $N \times N$ covariance matrix composed of blocks $C_{i,j}$. If $|a(i,j)|$ is small for many language pairs (as it is in the case of Wikipedia), then many empirical estimates $C_{i,j}$ are unreliable, which can result in overfitting. For this reason, we perform the truncated singular value decomposition on the matrix $C = [C_{1,2}  \cdots  C_{1,m}] \approx U S V^T$, where $U \in \RR^{n_1 \times k}, S \in \RR^{k \times k}, V \in \RR^{(\sum_{i=2}^m n_i) \times k}$. We split the matrix $V$ vertically in blocks with $n_2, \ldots, n_m$ rows: $V = [V_2^T  \cdots  V_m^T]^T$. Note that columns of $U$ are orthogonal but columns in each $V_i$ are not (columns of V are orthogonal). Let $V_1 := U$. We proceed by reducing the dimensionality of each $X_i$ by setting: $Y_i = V_i^T \cdot X_i$, where $Y_i \in \RR^{k\times N}$. To summarize, the first step reduces the dimensionality of the data and is based on CL-LSI, but optimizes only the hub language related cross-covariance blocks.

\paragraph{Step 2: Simplifying and solving SSCOR.}
The second step involves solving a generalized version of canonical correlation analysis on the matrices $Y_i$ in order to find the mappings $P_i$. The approach is based on the sum of squares of correlations formulation by Kettenring~\cite{Kettenring}, where we consider only correlations between pairs $(Y_1, Y_i), i >1$ due to the hub language problem characteristic.
We will present the original unconstrained optimization problem, then a constrained formulation based on the hub language problem characteristic. Then we will simplify the constraints and reformulate the problem as an eigenvalue problem by using Lagrange multipliers.

The original sum of squared correlations is formulated as an unconstrained problem:
\begin{equation*}
  \begin{aligned}
    & \underset{w_i \in \RR^{k}}{\text{maximize}}
    & & \sum_{i < j}^m  \rho(w_i^T Y_i, w_j^T Y_j)^2.
\end{aligned}
\end{equation*}
We solve a similar problem by restricting $i=1$ and omitting the optimization over non-hub language pairs.
Let $D_{i,i} \in \RR^{k \times k}$ denote the empirical covariance of $\mathcal{Y}_i$ and $D_{i,j}$ denote the empirical cross-covariance computed based on $\mathcal{Y}_i$ and $\mathcal{Y}_j$. We solve the following constrained (unit variance constraints) optimization problem:
\begin{equation}\label{squaredCorHubOriginal}
  \begin{aligned}
    & \underset{w_i \in \RR^{k}}{\text{maximize}}
    & & \sum_{i = 2}^m  \left(w_1^T D_{1,i} w_i \right)^2
    & \text{subject to}
    & & w_i^T D_{i,i} w_i = 1, \quad\forall i = 1,\ldots, m.
\end{aligned}
\end{equation}
The constraints $w_i^T D_{i,i} w_i$ can be simplified by using the Cholesky decomposition $D_{i,i} = K_i^T \cdot K_i$ and substitution: $y_i := K_i w_i$. By inverting the $K_i$ matrices and defining  $G_i := K_1^{-T} D_{1,i} K_i^{-1}$, the problem can be reformulated:
\begin{equation}\label{squaredCorHub}
  \begin{aligned}
    & \underset{y_i \in \RR^{k}}{\text{maximize}}
    & & \sum_{i = 2}^m  \left(y_1^T G_{i} y_i \right)^2
    & \text{subject to}
    & & y_i^T y_i = 1, \quad\forall i = 1,\ldots, m.
\end{aligned}
\end{equation}
A necessary condition for optimality is that the derivatives of the Lagrangian vanish. The Lagrangian of (\ref{squaredCorHub}) is expressed as:
$$  L(y_1, \ldots, y_m, \lambda_1, \ldots, \lambda_m) = \sum_{i = 2}^m  \left(y_1^T G_{i} y_i \right)^2 + \sum_{i=1}^m \lambda_i \left(y_i^T y_i - 1\right).$$
Stationarity conditions give us:
\begin{equation}\label{dLdx1}
 \frac{\partial}{\partial x_1} L = 0 \Rightarrow \sum_{i = 2}^m  \left(y_1^T G_{i} y_i \right) G_i y_i + \lambda_1 y_1 = 0,
\end{equation}
\begin{equation}\label{dLdxi}
\frac{\partial}{\partial x_i} L = 0 \Rightarrow \left(y_1^T G_{i} y_i \right) G_{i}^T y_1 + \lambda_i y_i = 0,~i > 1.
\end{equation}
Multiplying the equations (\ref{dLdxi}) with $y_i^T$ and applying the constraints, we can eliminate $\lambda_i$ which gives us:
\begin{equation}\label{eqy1yi}
G_{i}^T y_1 = \left(y_1^T G_{i} y_i \right) y_i,~i > 1.
\end{equation}
Plugging this into (\ref{dLdx1}), we obtain an eigenvalue problem:
$$\left( \sum_{i = 2}^m G_i G_{i}^T \right) y_1 + \lambda_1 y_1 = 0.$$
The eigenvectors of $\left( \sum_{i = 2}^m G_i G_{i}^T \right)$ solve the problem for the first language. The solutions for $y_i$ are obtained from (\ref{eqy1yi}): $y_i := \frac{G_{i}^T y_1}{\| G_{i}^T y_1 \|}$.
Note that the solution (\ref{squaredCorHubOriginal}) can be recovered by: $w_i := K_i^{-1} y_i$. The linear transformation of the $w$ variables are thus expressed as:
$$ Y_1 := \text{eigenvectors of} \sum_{i = 2}^m G_i G_{i}^T, $$
$$ W_1 = K_1^{-1} Y_1 $$
$$ W_i = K_i^{-1} G_{i}^T Y_1 N,$$
where $N$ is a diagonal matrix that normalizes $G_{i}^T Y_1$, with $N(j,j) := \frac{1}{\|G(_{i} Y_1(:,j)\|}$.

\paragraph{Remark.} The technique is related to  Generalization of Canonical Correlation Analysis (GCCA) by Carroll~\cite{Carroll}, where an unknown group configuration variable is defined and the objective is to maximize the sum of squared correlations between the group variable and the others. The problem can be reformulated as an eigenvalue problem. The difference lies in the fact that we set the unknown group configuration variable as the hub language, which simplifies the solution. The complexity of our method is $O(k^3)$, where $k$ is the reduced dimension from the LSI preprocessing step, whereas solving the GCCA method scales as $O(s^3)$, where $s$ is the number of samples (see~\cite{gifi}). Another issue with GCCA is that it cannot be directly applied to the case of missing documents.

To summarize, we first reduced the dimensionality of our data to $k$-dimensional features and then found a new representation (via linear transformation) that maximizes directions of linear dependence between the languages. The final projections that enable mappings to a common space are defined as: $P_i(x) = W_i^T V_i^T x.$

\section{Cross-lingual Event Linking}\label{sec:linking}

The main application on which we test the cross-lingual similarity is cross-lingual event linking. In online media streams -- particularly news articles -- there is often duplication of reporting, different viewpoints or opinions, all centering around a single event. The same events are covered by many articles and the question we address is how to find all the articles in different languages that are describing a single event. In this paper we consider the problem of matching events from different languages. We do not address the problem of detection of events and instead base our evaluation on an online system for detection of world events, Event Registry. The events are represented by clusters of articles and so ultimately our problem reduces to finding suitable matchings between clusters with articles in different languages.

\subsection{Problem definition}

The problem of cross-lingual event linking is to match monolingual clusters of news articles that describing the same event across languages. For example, we want to match a cluster of Spanish news articles and a cluster of English news articles that both describe the same earthquake.

Each article $a \in A$ is written in a language $\ell$, where $\ell \in L = \{\ell_1,\ell_2,...,\ell_m\}$. For each language $\ell$, we obtain a set of monolingual clusters $C_{\ell}$. More precisely, the articles corresponding to each cluster $c \in C_{\ell}$ are written in the language $\ell$. Given a pair of languages $\ell_a \in L$, $\ell_b \in L$ and $\ell_a \not= \ell_b$, we would like to identify all cluster pairs $(c_i, c_j) \in C_{\ell_a} \times C_{\ell_b}$ such that $c_i$ and $c_j$ describe the same event.

\begin{figure}[tb]
\centering
\includegraphics[width=0.7\textwidth]{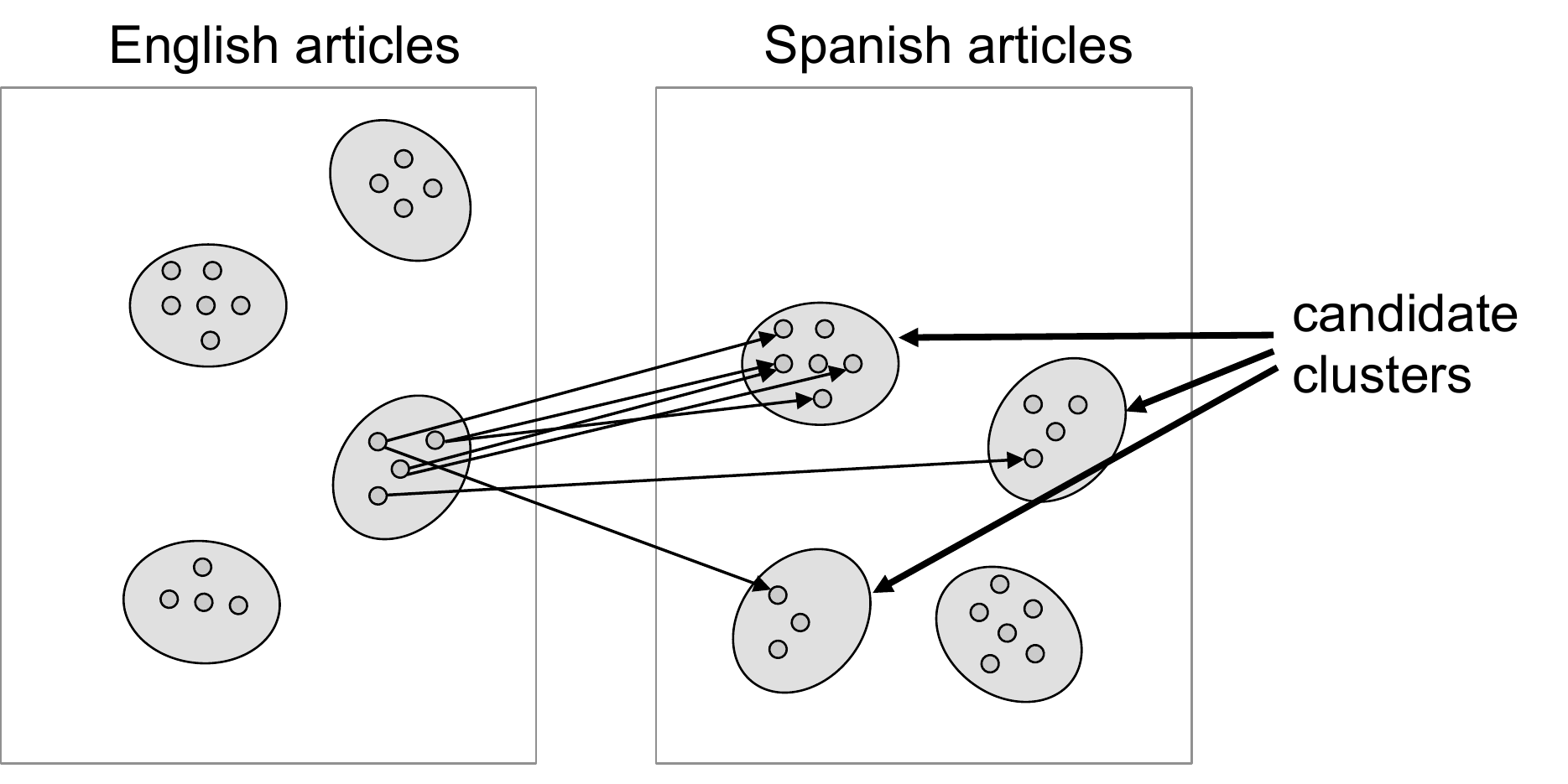}
\caption{\label{fig:clusters}  Clusters composed of English and Spanish news articles. Arrows link English articles with their Spanish $k$-nearest neighbor matches according to the cross-lingual similarity.}
\end{figure}

Matching of clusters is a \emph{generalized matching} problem. We cannot assume that there is only one cluster per language per event, nor can we assume complete coverage -- i.e., that there exists at least one cluster per event in every language. This is partly due to news coverage which might be more granular in some languages, partly due to noise and errors in the event detection process. This implies that we cannot make assumptions on the matching (e.g., one-to-one or complete matching) and excludes the use of standard weighted bipartite matching type of algorithms for this problem. An example is shown in Figure~\ref{fig:clusters}, where a cluster may contain articles which are closely matched with many clusters in a different language.

We are also seeking for an algorithm which does not do exhaustive comparison of all clusters, since that can become prohibitively expensive when working in a real-time setting. More specifically, we wish to avoid testing cluster $c_i$ with all the clusters from all the other languages. Performing exhaustive comparison would result in $O(|C|^2)$ tests, where $|C|$ is the number of all clusters (over all languages), which is not feasible when the number of clusters is on the order of tens of thousands. We address this by testing only clusters that are connected with at least one $k$-nearest neighbor (marked as \emph{candidate clusters} in Figure~\ref{fig:clusters}).

\subsection{Algorithm}\label{algo:features}

In order to identify clusters that are equivalent to cluster $c_i$, we have developed a two-stage algorithm. For a cluster $c_i$, we first efficiently identify a small set of candidate clusters and then find those clusters among the candidates, which are equivalent to $c_i$. An example is  shown in  Figure~\ref{fig:clusters}.

The details of the first step are described in Algorithm~\ref{cluster_merge_algo1}. The algorithm begins by individually inspecting each news article $a_i$ in the cluster $c_i$. Using a chosen method for computing cross-lingual document similarity (see Section~\ref{sec:models}), it identifies the 10 most similar news articles to $a_i$ in each language $\ell \in L$. For each similar article $a_j$, we identify its corresponding  cluster $c_j$ and add it to the set of candidates. The set of candidate clusters obtained in this way is several orders of magnitude smaller than the number of all clusters, and at most linear with respect to the number of news articles in cluster $c_i$. In practice, clusters contain highly related articles and as such similar articles from other languages mostly fall in only a few candidate clusters.

Although computed document similarities are approximate, our  assumption is that articles in different languages describing the same event will generally have a higher similarity than articles about different events. While this assumption does not always hold, redundancy in the data mitigates these false positives. Since we compute the 10 most similar articles for each article in $c_i$, we are likely to identify all the relevant candidates for cluster $c_i$.

\begin{algorithm}[t!]
\SetKwInput{KwInput}{input}\SetKwInput{KwOutput}{output}
\KwInput{test cluster $c_i$, a set of clusters $C_\ell$ for each language $\ell \in L$}
\KwOutput{a set of clusters $C$ that are potentially equivalent to $c_i$}
$C \leftarrow \{\}$\;
\For{article $a_i \in c_i$} {
    \For{language $\ell \in L$} {
        \tcc{use hub CCA to find 10 most similar articles to article $a_i$ in language $\ell$}
        $SimilarArticles = getCCASimilarArticles(a_i, \ell)$\;
        \For{article $a_j \in SimilarArticles$} {
            \tcc{find cluster $c_j$ to which article $a_j$ is assigned to}
            $c_j \leftarrow c$, such that $c \in C_\ell$ and $a_j \in c$\;
            \tcc{add cluster $c_j$ to the set of candidates $C$}
            $C \leftarrow C \cup \{ c_j \}$\;
        }
    }
}
\caption{Algorithm for identifying candidate clusters $C$ that are potentially equivalent to $c_i$}
\label{cluster_merge_algo1}
\end{algorithm}

The second stage of the algorithm determines which (if any) of the candidate clusters are equivalent to $c_i$. We treat this task as a supervised learning problem. For each candidate cluster $c_j \in C$, we compute a vector of learning features that should be indicative of whether $c_i$ and $c_j$ are equivalent or not and apply a binary classification model that predicts if the clusters are equivalent or not. The classification algorithm that we used to train a model was a linear Support Vector Machine (SVM) method~\cite{shawe-taylor04kernel}.

We use three groups of features to describe cluster pair $(c_i, c_j)$. The first group is based on {\bf cross-lingual article links}, which are derived using cross-lingual similarity: each news article $a_i$ is linked with its $10$-nearest neighbors articles from all other languages (10 per each language). The group contains the following features:

\begin{itemize}
\item \texttt{linkCount} is the number of times any of the news articles from $c_j$ is among $10$-nearest neighbors for articles from $c_i$. In other words, it is the number of times an article from $c_i$ has a very similar article (i.e., is among 10 most similar) in $c_j$.
\item \texttt{avgSimScore} is the average similarity score of the links, as identified for \texttt{linkCount}, between the two clusters.
\end{itemize}

The second group are {\bf concept-related features}. Articles that are imported into Event Registry are annotated by disambiguating mentioned \emph{entities} and \emph{keywords} to the corresponding Wikipedia pages~\cite{zhang2014}. Whenever Barack Obama is, for example, mentioned in the article, the article is annotated with a link to his Wikipedia page. In the same way, all mentions of entities (people, locations, organizations) and ordinary keywords (e.g., bank, tax, ebola, plane, company) are annotated. Although the Spanish article about Obama will be annotated with his Spanish version of the Wikipedia page, in many cases we can link the Wikipedia pages to their English versions. This can be done since Wikipedia itself provides information regarding which pages in different languages represent the same concept/entity. Using this approach, the word ``avi\'on'' in a Spanish article will be annotated with the same concept as the word ``plane'' in an English article. Although the articles are in different languages, the annotations can therefore provide a language-independent vocabulary that can be used to compare articles/clusters. By analyzing all the articles in clusters $c_i$ and $c_j$, we can identify the most relevant entities and keywords for each cluster. Additionally, we can also assign weights to the concepts based on how frequently they occur in the articles in the cluster. From the list of relevant concepts and corresponding weights, we consider the following features:

\begin{itemize}
\item \texttt{entityCosSim} is the cosine similarity between vectors of entities from clusters $c_i$ and $c_j$.
\item \texttt{keywordCosSim} is the cosine similarity between vectors of keywords from clusters $c_i$ and $c_j$.
\item \texttt{entityJaccardSim} is  Jaccard similarity coefficient~\cite{levandowsky1971} between sets of entities from clusters $c_i$ and $c_j$.
\item \texttt{keywordJaccardSim} is  Jaccard similarity coefficient between sets of keywords from clusters $c_i$ and $c_j$.
\end{itemize}

The last group of features contains three {\bf miscellaneous features} that seem discriminative but are unrelated to the previous two groups:
\begin{itemize}
\item \texttt{hasSameLocation} feature is a boolean variable that is true when the location of the event in both clusters is the same. The location of events is estimated by considering the locations mentioned in the articles that form a cluster and is provided by Event Registry.
\item \texttt{timeDiff} is the absolute difference in hours between the two events. The publication time and date of the events is computed as the average publication time and date of all the articles and is provided by Event Registry.
\item \texttt{sharedDates} is determined as the Jaccard similarity coefficient between sets of date mentiones extracted from articles. We use extracted mentions of dates provided by Event Registry, which uses an extensive set of regular expressions to detect and normalize mentions of dates in different forms.
\end{itemize}

\section{Evaluation}\label{sec:evaluation}

We will describe the main dataset for building cross-lingual models which is based on Wikipedia and then present two sets of experiments. The first set of experiments
establishes that the hub based approach can deal with language pairs where little or no training data is available. The second set of experiments compares the main approaches
that we presented on the task of mate retrieval and the task of event linking. Finally, we examine how different choices of features impact the event linking performance.

\subsection{Wikipedia Comparable Corpus}

To investigate the empirical performance of the low-rank approximations we will test the algorithms on a large-scale, real-world multilingual dataset that we extracted from Wikipedia by using inter-language links for alignment. This  results in a large number of weakly comparable documents in more than $200$ languages. Wikipedia is a large source of multilingual data that is especially important for the languages for which no translation tools, multilingual dictionaries as Eurovoc~\cite{eurovoc}, or strongly aligned multilingual corpora as Europarl~\cite{europarl} are available. Documents in different languages are related with so called 'inter-language' links that can be found on the left of the Wikipedia page. The Wikipedia is constantly growing. There are currently 12 Wikipedias with more than 1 million %$10^6$
 articles, $52$ with more than 100k %$10^5$
 articles, $129$ with more than 10k articles, and $236$ with more than $1,000$ articles.

Each Wikipedia page is embedded in the page tag. First, we check if the title of the page starts with a Wikipedia namespace (which includes categories and discussion pages) and do not process the page if it does. Then, we check if this is a redirection page and we store the redirect link because inter-language links can point to redirection links also. If none of the above applies, we extract the text and parse the Wikipedia markup. Currently, all the markup is removed.

We get inter-language link matrix using previously stored redirection links and inter-language links. If an inter-language link points to the redirection we replace it with the redirection target link. It turns out that we obtain the matrix $M$ that is not symmetric, consequently the underlying graph is not symmetric. That means that existence of the inter-language link in one way (i.e., English to German) does not guarantee that there is an inter-language link in the reverse direction (German to English). To correct this we transform this matrix to be symmetric by computing $M+M^T$ and obtaining an undirected graph. In the rare case that after symmetrization we have multiple links pointing from the document, we pick the first one that we encountered. This matrix enables us to build an alignment across all Wikipedia languages (for dumps available in 2013).

\subsection{Experiments With Missing Alignment Data}\label{experiments:hubcca}

 In this subsection, we will investigate the empirical performance of hub CCA approach. We will demonstrate that this approach can be successfully applied even in the case of fully missing alignment information.
 To this purpose, we select a subset of Wikipedia languages containing three major languages, English (4,212k articles)--\emph{en} (hub language), Spanish (9,686k articles)--\emph{es}, Russian (9,662k articles)--\emph{ru}, and five minority (in terms of Wikipedia sizes) languages, Slovenian (136k articles)--\emph{sl}, Piedmontese (59k articles)--\emph{pms}, Waray-Waray (112k articles)--\emph{war} (all with about 2 million native speakers), Creole (54k articles)--\emph{ht} (8 million native speakers), and Hindi (97k articles)--\emph{hi} (180 million native speakers). For preprocessing, we remove the documents that contain less than 20 different words (referred to as stubs\footnote{Such documents are typically of low value as a linguistic resource. Examples include the titles of the columns in the table, remains of the parsing process, or Wikipedia articles with very little or no information contained in one or two sentences.}) and remove words occurring in less than 50 documents as well as the top 100 most frequent words (in each language separately). We represent the documents as normalized TFIDF~\cite{Salton88term-weightingapproaches} weighted vectors. The IDF scores are computed for each language based on its aligned documents with the English Wikipedia. The English language IDF scores are based on all English documents for which aligned Spanish documents exist.

The evaluation is based on splitting the data into training and test sets. %(which are described later).
We select the test set documents as all multilingual documents with at least one nonempty alignment from the list: (\emph{hi}, \emph{ht}), (\emph{hi}, \emph{pms}), (\emph{war}, \emph{ht}), (\emph{war}, \emph{pms}). This guarantees that we cover all the languages. Moreover this test set is suitable for testing the retrieval through the hub as the chosen pairs have empty alignments. The remaining documents are used for training. In Table \ref{table:train_test}, we display the corresponding sizes of training and test documents for each language pair.

On the training set, we perform the two step procedure to obtain the common document representation as a set of mappings $P_i$. A test set for each language pair, $test_{i,j} = \{(x_\ell,y_\ell) | \ell = 1:n(i,j)\} $, consists of comparable document pairs (linked Wikipedia pages), where $n(i,j)$ is the test set size. We evaluate the representation by measuring mate retrieval quality on the test sets: for each $\ell$, we rank the projected documents $P_j(y_1),\ldots, P_j(y_{n(i,j)})$ according to their similarity with $P_i(x_\ell)$ and compute the rank of the mate document $r(\ell) = rank(P_j(y_\ell))$. The final retrieval score (between -100 and 100) is computed as: $\frac{100}{n(i,j)} \cdot \sum_{\ell = 1}^{n(i,j)} \left( \frac{n(i,j) - r(\ell)}{n(i,j) -1} -0.5\right)$. A score that is less than 0 means that the method performs worse than random retrieval and a score of 100 indicates perfect mate retrieval. The mate retrieval results are included in Table \ref{table:retrieval}.

We observe that the method performs well on all pairs of languages, where at least 50,000 training documents are available(\emph{en}, \emph{es}, \emph{ru}, \emph{sl}). We note that taking $k = 500$ or $k = 1,000$ multilingual topics usually results in similar performance, with some notable exceptions: in the case of (\emph{ht}, \emph{war}) the additional topics result in an increase in performance, as opposed to (\emph{ht}, \emph{pms}) where performance drops, which suggests overfitting. The languages where the method performs poorly are \emph{ht} and \emph{war}, which can be explained by the quality of data (see Table \ref{table:rank} and explanation that follows). In case of \emph{pms}, we demonstrate that solid performance can be achieved for language pairs (\emph{pms}, \emph{sl}) and (\emph{pms}, \emph{hi}), where only 2,000 training documents are shared between \emph{pms} and \emph{sl} and no training documents are available between \emph{pms} and \emph{hi}. Also observe that in the case of (\emph{pms}, \emph{ht}) the method still obtains a score of 62, even though training set intersection is zero and \emph{ht} data is corrupted, which we will show in the next paragraph.
{
\renewcommand\tabcolsep{3pt}
\begin{table}[h!]
\centering
\caption{Training -- test sizes (in thousands).
The first row represents the size of the training sets used to construct the mappings in low-dimensional language independent space using \emph{en} as a hub. The diagonal elements represent the number of the unique training documents and test documents in each language.
}
\label{table:train_test}
{
\small
\begin{tabular}{c|c|c|c|c|c|c|c|c|}
&	en&	es&	ru&	sl&	hi&	war&	ht&	pms\\\cline{1-9}
en&	671~-~4.64&	463~-~4.29&	369~-~3.19&	50.3~-~2&	14.4~-~2.76&	8.58~-~2.41&	 17~-~2.32&	16.6~-~2.67\\
\cline{2-9}
es&	\multicolumn{1}{c|}{}	&	463~-~4.29&	187~-~2.94&	28.2~-~1.96&	8.72~-~2.48&	 6.88~-~2.4&	13.2~-~2&	 13.8~-~2.58\\
\cline{3-9}
ru&	\multicolumn{2}{c|}{}	&	369~-~3.19&	29.6~-~1.92&	9.16~-~2.68&	2.92~-~1.1&	 3.23~-~2.2&	10.2~-~1.29\\
\cline{4-9}
sl&	\multicolumn{3}{c|}{}	&	50.3~-~2&	3.83~-~1.65&	1.23~-~0.986&	0.949~-~1.23&	 1.85~-~0.988\\
\cline{5-9}
hi&	\multicolumn{4}{c|}{}	&	14.4~-~2.76&	0.579~-~0.76&	0.0~-~2.08&	0.0~-~0.796\\
\cline{6-9}
war&	\multicolumn{5}{c|}{}	&	8.58~-~2.41&	0.043~-~0.534&	0.0~-~1.97\\
\cline{7-9}
ht&	\multicolumn{6}{c|}{}	&	17~-~2.32&	0.0~-~0.355\\
\cline{8-9}
pms&	\multicolumn{7}{c|}{}	&	16.6~-~2.67\\
\cline{9-9}
\end{tabular}
}
\end{table}
}

{
\renewcommand\tabcolsep{3pt}
\begin{table}[h!]
\caption{Pairwise retrieval, 500 topics on the left -- 1,000 topics on the right}\label{table:retrieval}
\begin{center}
\begin{tabular}{|c|c|c|c|c|c|c|c|c|}
\cline{1-9}
&	en&	es&	ru&	sl&	hi&	war&	ht&	pms\\\cline{1-9}
en&	    &	98~-~98&	95~-~97&	97~-~98&	82~-~84&	76~-~74&	53~-~55&	 96~-~97\\
\cline{1-9}
es&	97~-~98&	&	94~-~96&	97~-~98&	85~-~84&	76~-~77&	56~-~57&	96~-~96\\
\cline{1-9}
ru&	96~-~97&	94~-~95&	&	97~-~97&	81~-~82&	73~-~74&	55~-~56&	96~-~96\\
\cline{1-9}
sl&	96~-~97&	95~-~95&	95~-~95&	&	91~-~91&	68~-~68&	59~-~69&	93~-~93\\
\cline{1-9}
hi&	81~-~82&	82~-~81&	80~-~80&	91~-~91&	&	68~-~67&	50~-~55&	87~-~86\\
\cline{1-9}
war&	68~-~63&	71~-~68&	72~-~71&	68~-~68&	66~-~62&	&	28~-~48&	 24~-~21\\
\cline{1-9}
ht&	52~-~58&	63~-~66&	66~-~62&	61~-~71&	44~-~55&	16~-~50&	&	62~-~49\\
\cline{1-9}
pms&	95~-~96&	96~-~96&	94~-~94&	93~-~93&	85~-~85&	23~-~26&	66~-~54&	 \\
\cline{1-9}
\end{tabular}
\end{center}
\end{table}
}

We further inspect the properties of the training sets by roughly estimating the fraction $\frac{rank(A)}{min\left(rows\left(A\right),~cols\left(A\right)\right)}$ for each English training matrix and its corresponding mate matrix, where $rows(A)$ and $cols(A)$ denote the number of rows and columns respectively. The denominator represents the theoretically highest possible rank the matrix $A$ could have. Ideally, these two fractions should be approximately the same - both aligned spaces should have reasonably similar dimensionality. We display these numbers as pairs in Table \ref{table:rank}.

\begin{table}[h]
\caption{Dimensionality drift. Each column corresponds to a pair of aligned corpus matrices between English and another language. The numbers represent the ratio between the numerical rank and the highest possible rank. For example, the column $en -- ht$ tells us that for the English-Creole pairwise-aligned corpus matrix pair, the English counterpart has full rank, but the Creole counterpart is far having full rank.}
\label{table:rank}
\begin{tabular}{|c|c|c|c|c|c|c|}
\cline{1-7}
en -- es     &   en -- ru     &   en -- sl       &     en -- hi &   en -- war      &      en -- ht &   en -- pms\\
\cline{1-7}
0.81 -- 0.89   &  0.8 -- 0.89  &   0.98 -- 0.96    &    1 -- 1  &  0.74 -- 0.56  &      1 -- 0.22  &   0.89 -- 0.38\\
\cline{1-7}
\end{tabular}
\end{table}

It is clear that in the case of the Creole language only at most $22\%$ documents are unique and suitable for the training. Though we removed the stub documents, many of the remaining documents are nearly the same, as the quality of some smaller Wikipedias is low. This was confirmed for the Creole, Waray-Waray, and Piedmontese languages by manual inspection. The low quality documents correspond to templates about the year, person, town, etc. and contain very few unique words.

There is also a problem with the quality of the test data. For example, if we look at the test pair (\emph{war}, \emph{ht}) only 386/534 Waray-Waray test documents are unique but on the other side almost all Creole test documents (523/534) are unique. This indicates a poor alignment which leads to poor performance.
%}

\subsection{Evaluation Of Cross-Lingual Event Linking}
In order to determine how accurately we can predict cluster equivalence, we performed two experiments in a multilingual setting using English, German and Spanish languages for which we had labelled data to evaluate the linking performance. In the first experiment, we tested how well  the individual approaches for cross-lingual article linking perform when used for linking the clusters about the same event. In the second experiment we tested how accurate the prediction model is when trained on different subsets of learning features. To evaluate the prediction accuracy for a given dataset we used 10-fold cross validation.

We created a manually labelled dataset in order to evaluate cross-lingual event linking using two human annotators. The annotators were provided with an interface listing the articles, their content from and top concepts for a pair of clusters and their task was to determine if the clusters were equivalent or not (i.e., discuss same event). To obtain a pair of clusters $(c_i, c_j)$ to annotate, we first randomly chose a cluster $c_i$, used Algorithm~\ref{cluster_merge_algo1} to compute a set of potentially equivalent clusters $C$ and randomly chose a cluster $c_j \in C$. The dataset provided by the annotators contains 808 examples, of which 402 are equivalent clusters pairs and 406 are not. Clusters in each learning example are either in English, Spanish or German. Although Event Registry imports articles in other languages as well, we restricted our experiments to these three languages. We chose only these three languages since they have very large number of articles and clusters per day which makes the cluster linking problem hard due to large number of possible links.

In Section~\ref{sec:models}, we  described three main algorithms for identifying similar articles in different languages. These algorithms were $k$-means, LSI and hub CCA. As a training set, we used common Wikipedia alignment for all three languages. To test which of these algorithms performed best, we made the following test. For each of the three algorithms, we analyzed all articles in Event Registry and for each article computed the most similar articles in other languages. To test how informative the identified similar articles are for cluster linking we then trained three classifiers as described in Section~\ref{algo:features} -- one for each algorithm. Each classifier was allowed to use as learning features \textbf{only} the cross-lingual article linking features for which values are determined based on the selected algorithm ($k$-means, LSI and hub CCA). The results of the trained models are shown in Table~\ref{table:linkingEvalAlgos}. We also show how the number of topics (the dimensions of the latent space) influences the quality, except in the case of the $k$-means algorithm, where only the performance on 500 topic vectors is reported, due to higher computational cost.

We observe that, for the task of cluster linking, LSI and hub CCA perform comparably and both outperform $k$-means.

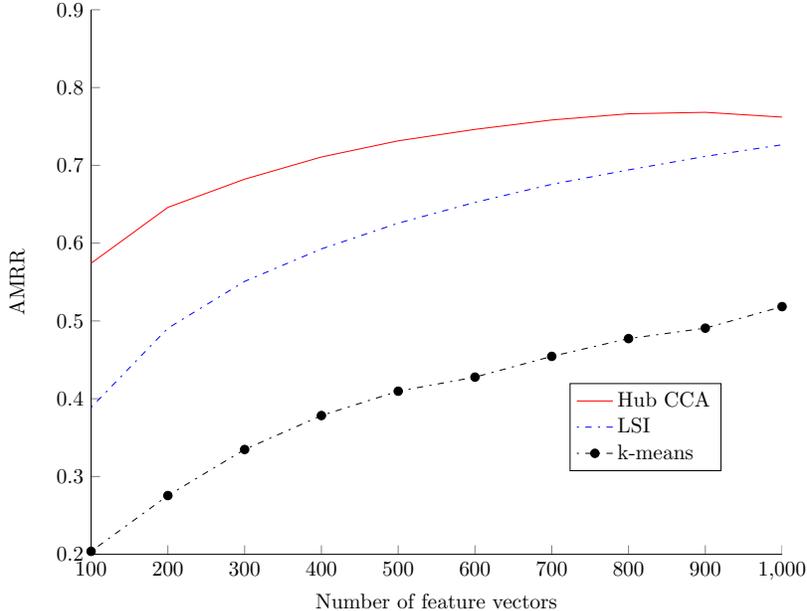
\begin{figure}
\centering
\begin{tikzpicture}[scale=0.8]

\begin{axis}[%
width=4.520833in,
height=3.565625in,
at={(0.758333in,0.48125in)},
scale only axis,
every outer x axis line/.append style={black},
every x tick label/.append style={font=\color{black}},
xmin=100,
xmax=1000,
xlabel={Number of feature vectors},
every outer y axis line/.append style={black},
every y tick label/.append style={font=\color{black}},
ymin=0.2,
ymax=0.9,
ylabel={AMRR},
axis x line*=bottom,
axis y line*=left,
legend style={at={(0.692559,0.152778)},anchor=south west,legend cell align=left,align=left,draw=black}
]
\addplot [color=red,solid]
  table[row sep=crcr]{%
100	0.574094231082746\\
200	0.645978450670063\\
300	0.682226506762222\\
400	0.710722094692085\\
500	0.731517169423203\\
600	0.746223870375363\\
700	0.758312886611436\\
800	0.766279310559898\\
900	0.768159613081689\\
1000	0.76200517888255\\
};
\addlegendentry{Hub CCA};

\addplot [color=blue,dash pattern=on 1pt off 3pt on 3pt off 3pt]
  table[row sep=crcr]{%
100	0.388773162433835\\
200	0.49061571487968\\
300	0.550801867794634\\
400	0.592565604079171\\
500	0.625657099289601\\
600	0.65240747960386\\
700	0.675602522126668\\
800	0.694149727717174\\
900	0.711511053167843\\
1000	0.726455461653153\\
};
\addlegendentry{LSI};

\addplot [color=black,dash pattern=on 1pt off 3pt on 3pt off 3pt,mark=*,mark options={solid}]
  table[row sep=crcr]{%
100	0.203799120889507\\
200	0.275537657758881\\
300	0.334738107118325\\
400	0.378380649062479\\
500	0.409668545430712\\
600	0.427765004442194\\
700	0.454449272479224\\
800	0.477291178317808\\
900	0.490692600634559\\
1000	0.518392823243787\\
};
\addlegendentry{k-means};

\end{axis}
\end{tikzpicture}%
\caption{Average of mean reciprocal ranks}
\label{pic:AMRR}
\end{figure}

% AMMR experiments
We also compared the proposed approaches on the task of Wikipedia mate retrieval (the same task as in Section~\ref{experiments:hubcca}). We computed the Average (over language pairs) Mean Reciprocal Rank (AMRR)~\cite{voorhees1999trec}  performance of the different approaches on the  Wikipedia data by holding out $15,000$ aligned test documents and using $300,000$ aligned documents as the training set. Figure~\ref{pic:AMRR} shows AMRR score as the function of the number of feature vectors. It is clear that hub CCA outperforms LSI approach and $k$-means lags far behind when testing on Wikipedia data. The hub CCA approach with $500$ topic vectors manages to perform comparably to the $LSI$-based approach with $1,000$ topic vectors, which shows that the $CCA$ method can improve both model memory footprint as well as similarity computation time.

% number of topics?
Furthermore, we inspected how the number of topics influences the accuracy of cluster linking. As we can see from Table~\ref{table:linkingEvalAlgos} choosing a number of features larger than $500$ barely affects linking performance, which is in contrast with the fact that additional topics helped to improve AMMR, see Figure~\ref{pic:AMRR}. Such differences may have arisen due to different domains of training and testing (Wikipedia pages versus news articles).

% cluster size?
We also analyzed how cluster size influences the accuracy of cluster linking. We would expect that if the tested pair of clusters has a larger number of articles then the classifier should be able to more accurately predict whether the clusters should be linked or not. The reasoning is that the large clusters would provide more document linking information (more articles mean more links to other similar articles) as well as more accurately aggregated semantic information. In the case of smaller clusters, the errors of the similarity models have greater impact which should decrease the performance of the classifier, too. To validate this hypothesis we have split the learning examples into two datasets -- one containing cluster pairs where the combined number of articles from both clusters is below 20 and one dataset where the combined number is 20 or more. The results of the experiment can be seen in Table~\ref{table:linkingEvalAlgosLargeSmall}. As it can be seen, the results confirm our expectations: for smaller clusters it is indeed harder to correctly predict if the cluster pair should be merged or not.

The hub CCA attains higher precision and classification accuracy on the task of linking small cluster pairs than the other methods, while LSI is slightly better on linking large cluster pairs. The gain in precision of LSI over hub CCA on linking large clusters is much smaller than the gain in precision of hub CCA over LSI on linking small clusters. For that reason we decided to use hub CCA as the similarity computation component in our system.

%
%%AMPR as function of the number of feature vectors
%%\includegraphics{slika.png}
%\centering
%\include{retrieval}
%\caption{Average of mean reciprocal ranks}
%\label{pic:AMRR}

\begin{table}[h]
\caption{Accuracy of cluster linking with 500/800/1,000 topic vectors obtained from different cross-lingual similarity algorithms. The table shows for each of the algorithms the obtained classification accuracy, precision and recall.}
\label{table:linkingEvalAlgos}
\begin{center}
\begin{tabular}{|c|c|c|c|c|}
  \hline
  \cline{1-5}
  Models & Accuracy \% & Precision \% & Recall \% & $F_1$ \% \\ \cline{1-5}
  hub CCA  & 78.2/79.6/80.3 & 76.3/78.0/80.5  & 81.6/82.1/79.9 & 78.9/80.0/80.2
  \\ \cline{1-5}
  LSI      & 78.9/78.7/80.6  & 76.8/77.0/78.7 & 83.3/80.6/83.6 & 79.9/78.8/81.1  \\ \cline{1-5}
 $k$-means & 73.9/-/- & 69.5/-/- & 84.6/-/- &  76.3/-/- \\ \cline{1-5}
 %$k$-means & 73.9/-/-\phantom{/78.7/80.6 } & 69.5/-/-\phantom{/78.7/80.6 }  & 84.6/-/-\phantom{/78.7/80.6 } &  76.3\phantom{/78.7/80.6 }  \\ \cline{1-5}
\end{tabular}
\end{center}
\end{table}

\begin{table}[h]
\caption{Accuracy of cluster linking using $500$ topic vectors on two datasets containing large (left number) and small (right number) clusters. The dataset with small clusters contained the subset of learning examples in which the combined number of articles from both clusters of the cluster pair were below 20. The remaining learning examples were put into the dataset of large clusters.}
\label{table:linkingEvalAlgosLargeSmall}
\begin{center}
\begin{tabular}{|c|c|c|c|c|}
  \hline
  \cline{1-5}
  Models & Accuracy \% & Precision \% & Recall \% & $F_1$ \% \\ \cline{1-5}
  hub CCA  & 81.2 - 77.8 & 80.5 - 74.5 & 91.3 - 57.5 & 85.6 - 64.9 \\ \cline{1-5}
  LSI      & 82.8 - 76.4 & 81.3 - 70.9 & 93.1 - 57.5 & 86.8 - 63.5 \\ \cline{1-5}
 $k$-means & 75.5 - 71.2 & 72.8 - 70.8 & 95.3 - 36.2 & 82.5 - 47.9 \\ \cline{1-5}
\end{tabular}
\end{center}
\end{table}

In the second experiment, we evaluate how relevant individual groups of features are to correctly determine cluster equivalence. For this purpose, we computed accuracy using individual groups of features, as well as using different combination of groups. Since hub CCA had the best  performance of the three algorithms, we used it to compute the values of the cross-lingual article linking features. The results of the evaluation are shown in Table~\ref{table:linkingEval}. We can see that using a single group of features, the highest prediction accuracy can be achieved using  concept-related features. The classification accuracy in this case is 88.5\%. By additionally including also the cross-lingual article linking features, the classification accuracy rises slightly to 89.4\%. Using all three groups of features, the achieved accuracy is 89.2\%.

To test if the accuracy of the predictions is language dependent we have also performed the evaluations separately on individual language pairs. For this experiment we have split the annotated learning examples into three datasets, where each dataset contained only examples for one language pair. When training the classifier all three groups of features were available. The results are shown in Table~\ref{table:langPairEval}. We can see that the performance of cluster linking on the English-German dataset is the highest in terms of accuracy, precision, recall and $F_1$. The performance on the English-Spanish dataset is comparable to the performance on the English-German dataset, where the former achieves higher recall (and slightly higher $F_1$ score), while the latter achieves higher precision. A possible explanation of these results is that the higher quantity and quality of English-German language resources leads to a more accurate cross-lingual article similarity measure as well as to a more extensive semantic annotation of the articles.

Based on the performed experiments, we can make the following conclusions. The cross-lingual similarity algorithms provide valuable information that can be used to identify clusters that describe the same event in different languages. For the task of cluster linking, the cross-lingual article linking features are however significantly less informative compared to the concept-related features that are extracted from the semantic annotations. Nevertheless, the cross-lingual article similarity features are very important for two reasons. The first  is that they allow us to identify for a given cluster a limited set of candidate clusters that are potentially equivalent. This is a very important feature since it reduces the search space by several orders of magnitude. The second reason these features are important is that concept annotations are not available for all articles as the annotation of news articles is computationally intensive and can only be done for a subset of collected articles. The prediction accuracies for individual language pairs are comparable although it seems that the achievable accuracy correlates with the amount of available language resources.

\begin{table}[h]
\caption{The accuracy of the classifier for story linking using different sets of learning features.}
\label{table:linkingEval}
\begin{center}
\begin{tabular}{|c|c|c|c|c|}
  \hline
  \cline{1-5}
  Features & Accuracy \% & Precision \% & Recall \% & $F_1$ \%  \\ \cline{1-5}
  hub CCA            & $78.3 \pm 5.9$ & $78.2 \pm  7.0$ & $78.9 \pm  5.2$ & $78.4 \pm  5.5$ \\ \cline{1-5}
  Concepts           & $88.5 \pm 2.7$ & $88.6 \pm  4.8$ & $88.6 \pm  2.2$ & $88.5 \pm  2.4$ \\ \cline{1-5}
  Misc               & $54.8 \pm 6.7$ & $61.8 \pm 16.5$ & $58.2 \pm 30.2$ & $52.4 \pm 13.0$ \\ \cline{1-5}
  hub CCA + Concepts & $89.4 \pm 2.5$ & $89.4 \pm  4.6$ & $89.6 \pm  2.4$ & $89.4 \pm  2.3$ \\ \cline{1-5}
  hub CCA + Misc     & $78.8 \pm 5.0$ & $78.9 \pm  7.1$ & $79.4 \pm  4.6$ & $79.0 \pm  4.5$ \\ \cline{1-5}
  Concepts + Misc    & $88.7 \pm 2.6$ & $88.8 \pm  4.6$ & $88.8 \pm  2.2$ & $88.7 \pm  2.3$ \\ \cline{1-5}
  All                & $89.2 \pm 2.6$ & $88.8 \pm  4.9$ & $90.1 \pm  1.9$ & $89.3 \pm  2.3$ \\ \cline{1-5}
  \hline
\end{tabular}
\end{center}
\end{table}

\begin{table}[h]
\caption{The accuracy of the classifier for story linking on training data for each language pair separately using all learning features.}
\label{table:langPairEval}
\begin{center}
\begin{tabular}{|c|c|c|c|c|}
  \hline
  \cline{1-5}
  Language pair & Accuracy \% & Precision \% & Recall \% & $F_1$ \% \\ \cline{1-5}
  en, de & $91.8 \pm 5.5$ & $91.7 \pm  6.3$ & $93.7 \pm  6.3$ & $92.5 \pm  5.1$ \\ \cline{1-5}
  en, es & $87.7 \pm 5.4$ & $87.7 \pm  7.4$ & $88.5 \pm  9.8$ & $87.6 \pm  5.9$ \\ \cline{1-5}
  es, de & $88.6 \pm 4.3$ & $89.7 \pm  9.1$ & $84.3 \pm 11.9$ & $85.9 \pm  6.0$ \\ \cline{1-5}
  \hline
\end{tabular}
\end{center}
\end{table}

\subsection{Remarks on the scalability of the implementation}

One of the main advantages of our approach is that it is highly scalable. It is fast, very robust to quality of training data, easily extendable, simple to implement and has relatively small hardware requirements. The similarity pipeline is the most computationally intensive part and currently runs on a machine with two Intel Xeon E5-2667 v2, 3.30GHz processors with 256GB of RAM. This is sufficient to do similarity computation over a large number of languages if needed. It currently uses Wikipedia as a freely available knowledge base and experiments show that the similarity pipeline dramatically reduces the search space when linking clusters.

Currently, we compute similarities over $24$ languages with tags: \emph{eng}, \emph{spa}, \emph{deu}, \emph{zho}, \emph{ita}, \emph{fra}, \emph{rus}, \emph{swe}, \emph{nld}, \emph{tur}, \emph{jpn}, \emph{por}, \emph{ara}, \emph{fin}, \emph{ron}, \emph{kor}, \emph{hrv}, \emph{tam}, \emph{hun}, \emph{slv}, \emph{pol}, \emph{srp}, \emph{cat}, \emph{ukr} but we support any language from the top $100$ Wikipedia languages. Our data stream is Newsfeed (\protect\url{http://newsfeed.ijs.si/}) which provides 430k unique articles per day. Our system currently computes 2 million similarities per second, that means that we compute $16 \cdot 10^{10}$ similarities per day. We
store one day buffer for each language which requires 1.5 GB of memory with documents   stored as 500-dimensional vectors. We  note that the time complexity of the similarity computations scales linearly with dimension of the feature space and does not  depend on number of languages. For each article, we compute the top $10$  most similar ones in every other language.

For all linear algebra matrix and vector operations, we use high performance numerical linear algebra libraries as BLAS, OPENBLAS and Intel MKL, which currently allows us to process more than one million articles per day.
In our current implementation, we use the variation of the hub approach. Our projector matrices are of size $500\times 300,000,$ so every projector takes about $1.1$ GB of RAM. Moreover, we need proxy matrices of size $500\times500$ for every language pair. That is 0.5 GB for $24$ languages and $9.2$ GB for $100$ languages. All together we need around 135 GB of RAM for the system with 100 languages.
 Usage of proxy matrices enables the projection of all input documents in the common space and handling language pairs with missing or low alignment. That enables us to do block-wise similarity computations further improving system efficiency. Our code can therefore be easily parallelized using matrix multiplication rather than performing more matrix - vector multiplications. This speeds up our code by a factor of around $4.$ In this way, we obtain some caching gains and ability to use vectorization.
Our system is also easily extendable. Adding a new language requires the  computation of  a projector matrix and proxy matrices with all other already available languages.

\section{Discussion and Future Work}

In this paper we have presented a cross-lingual system for linking events in different languages. Building on an existing system, Event Registry, we present and evaluate several approaches to compute a  cross-lingual similarity function. We also present an approach to link events and evaluate effectiveness of various features.  The  final pipeline is scalable both in terms of number of articles and number of languages, while accurately linking events.

On the task of mate retrieval, we observe that refining the LSI-based projections with hub CCA leads to improved retrieval precision, but the methods perform comparably on the task of event linking. Further inspection showed that the CCA-based approach reached a higher precision on smaller clusters. The interpretation is that the linking features are highly aggregated for large clusters, which compensates the lower per-document precision of LSI. Another possible reason is that the advantage that we show on Wikipedia is lost on the news domain. This hypothesis could be validated by testing the approach on documents from a different domain.

The experiments show that the hub CCA-based features present a good baseline, which can greatly benefit from additional semantic-based features. Even though in our experiments the addition of CCA-based features to semantic features did not lead to great performance improvements, there are two important benefits in the approach. First, the linking process can be sped up by using a smaller set of candidate clusters. Second, the approach is robust to languages where semantic extraction is not available, due to scarce linguistic resources.

\subsection{Future Work}

Currently the system is loosely-coupled -- the language component is built independently from the rest of the system, in particular the linking component. It is possible that better embeddings can be obtained by methods that jointly optimize a classification task and the embedding.
		
Another point of interest is to evaluate the system on languages with scarce linguistic resources, where semantic annotation might not be available. For this purpose, the labelled dataset of linked clusters should be extended first. The mate retrieval evaluation showed that even for language pairs with no training set overlap, the hub CCA recovers some signal.

In order to further improve the performance of the classifier for cluster linking, additional features should also be extracted from articles and clusters and checked if they can increase the accuracy of the classification. Since the amount of linguistic resources vary significantly from language to language it would also make sense to build a separate classifier for each language pair. Intuitively, this should improve performance since weights of individual learning features could be adapted to the tested pair of languages.

%\acks{The authors wish to thank blah blah blah.}


\begin{thebibliography}{10}

\bibitem{brank2014}
Janez Brank, Gregor Leban, and Marko Grobelnik.
\newblock A high-performance multithreaded approach for clustering a stream of
  documents.
\newblock In {\em Proceedings of the 17th International Multiconference
  Information Society 2014, Volume E, Ljubljana, Slovenia}, pages 5--8, 2014.

\bibitem{Carroll}
J.~D. Carroll.
\newblock Generalization of canonical correlation analysis to three or more
  sets of variables.
\newblock {\em Proceedings of the American Psychological Association}, pages
  227--228, 1968.

\bibitem{lsi}
S.~Deerwester, S.~T. Dumais, T.~K. Landauer, G.~W. Furnas, and R.~A. Harshman.
\newblock Indexing by latent semantic analysis.
\newblock {\em Journal of the American Society for Information Science},
  41(6):391--407, 1990.

\bibitem{cl_lsi}
S.T. Dumais, T.A. Letsche, M.L. Littman, and T.K Landauer.
\newblock Automatic cross-language retrieval using latent semantic indexing.
\newblock In {\em AAAI spring symposium on cross-language text and speech
  retrieval. American Association for Artificial Intelligence, vol. 16. 1997},
  page~21, 1997.

\bibitem{mrpqr}
B.~Fortuna, N.~Cristianini, and J.~Shawe-Taylor.
\newblock {\em Kernel methods in bioengineering, communications and image
  processing}, chapter A Kernel Canonical Correlation Analysis For Learning The
  Semantics Of Text, pages 263--282.
\newblock Idea Group Publishing, 2006.

\bibitem{gifi}
Albert Gifi.
\newblock {\em Nonlinear Multivariate Analysis}.
\newblock Wiley Series in Probability and Statistics, 1990.

\bibitem{golub}
Gene~H Golub and Charles~F Van~Loan.
\newblock {\em Matrix computations}, volume~3.
\newblock Johns Hopkins University Press, 2012.

\bibitem{tropp}
N.~Halko, P.~G. Martinsson, and J.~A. Tropp.
\newblock Finding structure with randomness: Probabilistic algorithms for
  constructing approximate matrix decompositions.
\newblock {\em Society for Industrial and Applied Mathematics Review},
  53(2):217--288, May 2011.

\bibitem{Hardoon_usingimage}
David~R Hardoon, Janaina Mourao-Miranda, Michael Brammer, and John
  Shawe-Taylor.
\newblock Using image stimuli to drive fmri analysis.
\newblock In {\em Neural Information Processing}, pages 477--486. Springer,
  2008.

\bibitem{kmeans}
John Hartigan.
\newblock {\em Clustering algorithms}.
\newblock John Wiley \& Sons Inc, New York, 1975.

\bibitem{Hotelling}
Harold Hotelling.
\newblock The most predictable criterion.
\newblock {\em Journal of educational Psychology}, 26(2):139, 1935.

\bibitem{Kettenring}
J.~R. Kettenring.
\newblock Canonical analysis of several sets of variables.
\newblock {\em Biometrika}, 58:433--45, 1971.

\bibitem{europarl}
Philipp Koehn.
\newblock Europarl: A parallel corpus for statistical machine translation.
\newblock In {\em The Tenth Machine Translation Summit}, volume~5, pages
  79--86, Phuket, Thailand, 2005.

\bibitem{moses}
Philipp Koehn, Hieu Hoang, Alexandra Birch, Chris Callison-Burch, Marcello
  Federico, Nicola Bertoldi, Brooke Cowan, Wade Shen, Christine Moran, Richard
  Zens, Chris Dyer, Ond\v{r}ej Bojar, Alexandra Constantin, and Evan Herbst.
\newblock Moses: Open source toolkit for statistical machine translation.
\newblock In {\em Proceedings of the 45th Annual Meeting of the ACL on
  Interactive Poster and Demonstration Sessions}, ACL '07, pages 177--180,
  Prague, Czech Republic, 2007. Association for Computational Linguistics.

\bibitem{Leban2014I}
Gregor Leban, Blaz Fortuna, Janez Brank, and Marko Grobelnik.
\newblock Cross-lingual detection of world events from news articles.
\newblock In {\em Proceedings of the 13th International Semantic Web
  Conference}, pages 21--24, Riva del Garda - Trentino, Italy, 2014.

\bibitem{Leban2014W}
Gregor Leban, Blaz Fortuna, Janez Brank, and Marko Grobelnik.
\newblock Event registry: Learning about world events from news.
\newblock In {\em Proceedings of the Companion Publication of the 23rd
  International Conference on World Wide Web Companion}, WWW Companion '14,
  pages 107--110, Seoul, Republic of Korea, 2014. International World Wide Web
  Conferences Steering Committee.

\bibitem{Leetaru2013Gdelt}
Kalev Leetaru and Philip~A Schrodt.
\newblock Gdelt: Global data on events, location, and tone, 1979--2012.
\newblock In {\em International Studies Association (ISA) Annual Convention},
  volume~2, page~4, San Francisco, California, USA, 2013.

\bibitem{levandowsky1971}
Michael Levandowsky and David Winter.
\newblock Distance between sets.
\newblock {\em Nature}, 234(5323):34--35, 11 1971.

\bibitem{Milne:2008:LLW:1458082.1458150}
David Milne and Ian~H. Witten.
\newblock Learning to link with wikipedia.
\newblock In {\em Proceedings of the 17th ACM Conference on Information and
  Knowledge Management}, CIKM '08, pages 509--518, Napa Valley, California,
  USA, 2008. ACM.

\bibitem{polyLDA}
David Mimno, Hanna~M. Wallach, Jason Naradowsky, David~A. Smith, and Andrew
  McCallum.
\newblock Polylingual topic models.
\newblock In {\em Proceedings of the 2009 Conference on Empirical Methods in
  Natural Language Processing: Volume 2 - Volume 2}, EMNLP '09, pages 880--889,
  Stroudsburg, PA, USA, 2009. Association for Computational Linguistics.

\bibitem{iti}
Andrej Muhi{\v{c}}, Jan Rupnik, and Primo{\v{z}} {\v{S}}kraba.
\newblock Cross-lingual document similarity.
\newblock In {\em Information Technology Interfaces (ITI), Proceedings of the
  ITI 2012 34th International Conference on}, pages 387--392, Cavtat /
  Dubrovnik, Croatia, 2012. IEEE.

\bibitem{Pearson1901On}
K.~Pearson.
\newblock {On lines and planes of closest fit to systems of points in space}.
\newblock {\em Philosophical Magazine}, 2(6):559--572, 1901.

\bibitem{multilingualBook}
Carol Peters and Martin Braschler.
\newblock {\em Multilingual Information Retrieval}.
\newblock Springer Berlin Heidelberg, Berlin, Heidelberg, 2012.

\bibitem{platt2010translingual}
John~C Platt, Kristina Toutanova, and Wen-tau Yih.
\newblock Translingual document representations from discriminative
  projections.
\newblock In {\em Proceedings of the 2010 Conference on Empirical Methods in
  Natural Language Processing}, pages 251--261, Massachusetts, USA, 2010.
  Association for Computational Linguistics.

\bibitem{plagiarism}
Martin Potthast, Alberto Barr\'{o}n-Cede\~{n}o, Benno Stein, and Paolo Rosso.
\newblock Cross-language plagiarism detection.
\newblock {\em Language Resources and Evaluation}, 45(1):45--62, March 2011.

\bibitem{ESA}
Martin Potthast, Benno Stein, and Maik Anderka.
\newblock A wikipedia-based multilingual retrieval model.
\newblock In {\em Advances in Information Retrieval , 30th European Conference
  on Information Retrieval Research (ECIR)}, pages 522--530, Glasgow, UK, 2008.

\bibitem{pouliquen2008story}
Bruno Pouliquen, Ralf Steinberger, and Olivier Deguernel.
\newblock Story tracking: linking similar news over time and across languages.
\newblock In {\em Proceedings of the Workshop on Multi-source Multilingual
  Information Extraction and Summarization}, pages 49--56, Manchester, United
  Kingdom, 2008. Association for Computational Linguistics.

\bibitem{Pouliquen2006}
Bruno Pouliquen, Ralf Steinberger, and Camelia Ignat.
\newblock Automatic annotation of multilingual text collections with a
  conceptual thesaurus.
\newblock {\em arXiv preprint cs/0609059}, 2006.

\bibitem{eurovoc}
Jose Mar\'{i}a~\'{A}lvarez Rodr\'{i}guez, Emilio~Rubiera Azcona, and Luis~Polo
  Paredes.
\newblock Promoting government controlled vocabularies for the semantic web:
  the eurovoc thesaurus and the cpv product classification system.
\newblock {\em Semantic Interoperability in the European Digital Library}, page
  111, 2008.

\bibitem{nips}
Jan Rupnik, Andrej Muhic, and Primoz Skraba.
\newblock Low-rank approximations for large, multi-lingual data.
\newblock {\em Low Rank Approximation and Sparse Representation, Neural
  Information Processing Systems 2011 Workshop}, 2011.

\bibitem{nips2}
Jan Rupnik, Andrej Muhic, and Primoz Skraba.
\newblock Spanning spaces: Learning cross-lingual similarities.
\newblock {\em Beyond Mahalanobis: Supervised Large-Scale Learning of
  Similarity, Neural Information Processing Systems 2011 Workshop}, 2011.

\bibitem{sikdd}
Jan Rupnik, Andrej Muhic, and Primoz Skraba.
\newblock Multilingual document retrieval through hub languages.
\newblock In {\em Proceedings of the 15th Multiconference on Information
  Society 2012 (IS-2012)}, pages 201--204, Ljubljana, Slovenia, 2012.

\bibitem{Salton88term-weightingapproaches}
Gerard Salton and Christopher Buckley.
\newblock Term-weighting approaches in automatic text retrieval.
\newblock volume~24, pages 513--523. Elsevier, 1988.

\bibitem{shawe-taylor04kernel}
John Shawe-Taylor and Nello Cristianini.
\newblock {\em Kernel Methods for Pattern Analysis}.
\newblock Cambridge University Press, 2004.

\bibitem{Steinberger2008}
Ralf Steinberger, Bruno Pouliquen, and Camelia Ignat.
\newblock Newsexplorer: Multilingual news analysis with cross-lingual linking.
\newblock {\em Information Technology Interfaces}, 2005.

\bibitem{Trampus2012}
Mitja Trampu\v{s} and Bla\v{z} Novak.
\newblock The internals of an aggregated web news feed.
\newblock In {\em Proceedings of 15th Multiconference on Information Society
  2012 (IS-2012)}, pages 221--224, Ljubljana, Slovenia, 2012.

\bibitem{voorhees1999trec}
Ellen~M Voorhees et~al.
\newblock The trec-8 question answering track report.
\newblock In {\em Proceedings of the 8th Text Retrieval Conference (TREC-8)},
  volume~99, pages 77--82, Gaithersburg, MD, USA, 1999.

\bibitem{nonnegfactor_lsi}
Min Xiao and Yuhong Guo.
\newblock A novel two-step method for cross language representation learning.
\newblock In {\em Advances in Neural Information Processing Systems}, pages
  1259--1267, Sateline, NV, USA, 2013.

\bibitem{PCL_LSA}
Duo Zhang, Qiaozhu Mei, and ChengXiang Zhai.
\newblock Cross-lingual latent topic extraction.
\newblock In {\em Proceedings of the 48th Annual Meeting of the Association for
  Computational Linguistics}, pages 1128--1137, Uppsala, Sweden, 2010.
  Association for Computational Linguistics.

\bibitem{zhang2014saaacamactat}
Lei Zhang and Achim Rettinger.
\newblock Semantic annotation, analysis and comparison: A multilingual and
  cross-lingual text analytics toolkit.
\newblock In {\em Proceedings of the Demonstrations at the 14th Conference of
  the European Chapter of the Association for Computational Linguistics (EACL
  2014)}, pages 13--16, Gothenburg, Sweden, April 2014. Association for
  Computational Linguistics.

\bibitem{zhang2014}
Lei Zhang and Achim Rettinger.
\newblock X-lisa: Cross-lingual semantic annotation.
\newblock {\em Proceedings of the Very Large Data Bases (VLDB) Endowment},
  7(13):1693--1696, August 2014.

\end{thebibliography}
\end{document}